\PassOptionsToPackage{hyphens}{url}
\documentclass[
  aps,
  prx,             
  reprint,           
  superscriptaddress,
  longbibliography,  
  floatfix
]{revtex4-2}

\usepackage[T1]{fontenc}
\usepackage[utf8]{inputenc}

\usepackage{amsmath,amssymb,amsfonts}
\usepackage{physics}
\usepackage{braket}
\usepackage{graphicx}
\usepackage{xcolor}
\usepackage{hyperref}
\usepackage{cleveref}
\usepackage{booktabs}

\usepackage[caption=false]{subfig}

\usepackage{multirow}
\usepackage{siunitx}
\usepackage[ruled,vlined,linesnumbered]{algorithm2e}

\usepackage{tikz}
\usepackage{quantikz}
\usetikzlibrary{quantikz2}
\usepackage{todonotes}
\usepackage{enumitem}

\begin{document}

\title{Error Mitigation of Fault-Tolerant Quantum Circuits with Soft Information}

\author{Zeyuan Zhou}
\email{v.zhou@yale.edu}
\affiliation{Department of Computer Science, Yale University, New Haven, CT, USA}
\affiliation{Yale Quantum Institute, Yale University, New Haven, CT, USA}

\author{Shaun Pexton}
\email{shaun.pexton@yale.edu}
\affiliation{Department of Computer Science, Yale University, New Haven, CT, USA}
\affiliation{Department of Applied Physics, Yale University, New Haven, CT, USA}

\author{Aleksander Kubica}
\email{a.kubica@yale.edu}
\affiliation{Yale Quantum Institute, Yale University, New Haven, CT, USA}
\affiliation{Department of Applied Physics, Yale University, New Haven, CT, USA}

\author{Yongshan Ding}
\email{yongshan.ding@yale.edu}
\affiliation{Department of Computer Science, Yale University, New Haven, CT, USA}
\affiliation{Yale Quantum Institute, Yale University, New Haven, CT, USA}
\affiliation{Department of Applied Physics, Yale University, New Haven, CT, USA}

\begin{abstract}
Quantum error mitigation (QEM) is typically viewed as a suite of practical techniques for today’s noisy intermediate-scale quantum devices, with limited relevance once fault-tolerant quantum computers become available.
In this work, we challenge this conventional wisdom by showing that QEM can continue to provide substantial benefits in the era of quantum error correction (QEC) and in an even more efficient manner than it does on current devices.
We introduce a framework for logical-level QEM that leverages soft information naturally produced by QEC decoders, requiring no additional data, hardware modifications, or runtime overhead beyond what QEC protocols already provide.
Within this framework, we develop and analyze three logical-level QEM techniques: post-selection and runtime abort policy, probabilistic error cancellation, and zero-noise extrapolation.
Our techniques can reduce logical error rates by more than 100× while discarding fewer than 0.1\% of shots; they also provide \textit{in situ} characterization of logical channels for QEM protocols.
As a proof of principle, we benchmark our approach using a surface-code architecture and two state-of-the-art decoders based on tensor-network contraction and minimum-weight perfect matching.
We evaluate the performance of logical-level QEM on random Clifford circuits and molecular simulation algorithms and find that, compared to previous approaches relying on QEC only or QEC combined with QEM, we can achieve up to 87.4\% spacetime overhead savings.
Our results demonstrate that logical-level QEM with QEC decoder soft information can reliably improve logical performance, underscoring the efficiency and usefulness of QEM techniques for fault-tolerant quantum computers.
\end{abstract}

\maketitle

% ---- body ----
\section{Introduction}
\label{sec:introduction}
Quantum error-correcting (QEC) codes have emerged as one of the most important building blocks for 
fault-tolerant quantum computation (FTQC) ~\cite{Shor1995decoherence,Steane1996,Gottesman1998FTtheory,Preskill1998qc,shor1996fault,Dennis2002topological}.
However, realizing high-distance QEC codes is a complex, resource-intensive experimental and engineering task.
First of all, prevailing hardware platforms have 
connectivity and control constraints, making long-range stabilizer checks difficult to implement 
\cite{wu2022synthesis, Bravyi_2024LDPCBB}. Secondly, large-scale real-time decoding is computationally demanding, requiring 
both high throughput and low latency \cite{Bausch2023MLdecoding, wu2023fusionblossomfastmwpm}. 
Thirdly, realistic noise models (that include coherent errors, crosstalk, etc.) are beyond the typical QEC assumptions 
\cite{Huang2019coherentqec,Zhou2023crosstalkctrl, Beale2018decoheres, Iverson2020coherenceinlogical, 
katsuda2024qecrealistic, zhou2025surfacecodecrosstalk} and pose challenges to the actual code performance.

To take full advantage of today's noisy intermediate-scale quantum devices incapable of realizing high-distance QEC codes, ingenious quantum error mitigation (QEM) protocols have been introduced~\cite{Temme2017QEM}, demonstrating utility in a wide range of practical algorithms~\cite{Kandala_2019QEMextends, Kim2023utility, zhang2025demonstratingquantumerrormitigation, Wahl2023ZNEscaling}. 
Common QEM techniques include probabilistic error cancellation (PEC) \cite{Temme2017QEM, PEC_lindbladian2023Ewout, PECscaling2021Andrea, PECdynamic2024Gupta, SimPEC2023Guimaraes, koh2025readouterrormitigationmidcircuit} and zero-noise extrapolation (ZNE) 
\cite{Temme2017QEM, Giurgica_Tiron_2020ZNE, He2020ZNEinsert, Majumdar2023bestQEM}. For both techniques, a critical subroutine is noise characterization. The goal is to 
accurately estimate the channel of each noisy gate and then use the information to either invert the channel or amplify the noise for extrapolation.

Besides QEM, noise characterization itself is a very important procedure for calibration and validation of physical qubits and gates.
Existing protocols include gate-set tomography 
(GST) \cite{Nielsen_2021GST, Brieger2023CGST}, quantum process tomography \cite{Huang_2020Classicalshadow}, 
randomized benchmarking \cite{Knill_2008RB,Hines2023RBdemo}, and cycle benchmarking \cite{Zhang2025cb}. 
Those self-consistent learning protocols all incur high shot overhead that is infeasible for large-scale quantum algorithms, 
and are inaccurate in the presence of noise drifting \cite{wang2024drift, Proctor2020drift}.

In this paper, we challenge the conventional belief that QEM techniques are primarily useful for today's noisy quantum devices and of limited relevance in the era of QEC and FTQC.
We provide theoretical foundations for logical-level QEM and end-to-end protocols leveraging soft information provided by QEC decoders. 
Importantly, there is no need for any additional characterization experiments, as classical syndrome information gathered during QEC cycles suffices for unbiased estimation of the logical channel and logical-level QEM.
The decoder soft information is per-gate addressable, meaning that it can characterize any logical gate.
Our protocol is also per-shot addressable, meaning that each shot returns not just a Bernoulli outcome but the posterior probability that reflects a non-drifted noise process during the execution of the algorithm. 
The estimator also has a better variance 
scaling compared to the existing estimators.

\begin{figure}[!t]
    \centering
    \includegraphics[width=\linewidth]{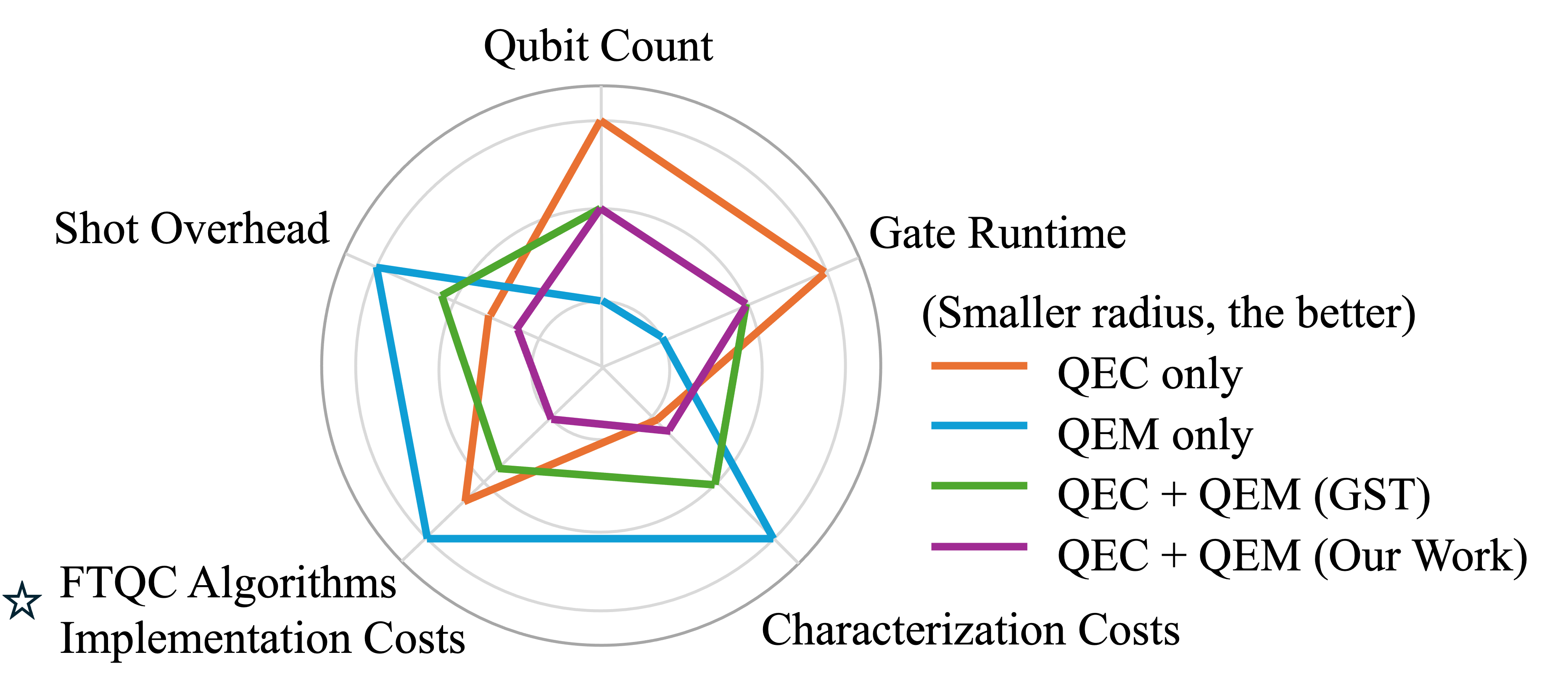}
    \caption{
    Qualitative resource overhead map of the four architectures to implement FTQC algorithms: QEC only, QEM only, GST-based QEC+QEM, and soft information-based QEC+QEM. A larger radial coordinate indicates a higher resource overhead. Our soft information-based QEC+QEM reduces demands in multiple dimensions, leading to a much more feasible architecture for fault tolerance.}
    \label{fig:overview}
\end{figure}

In the context of FTQC with the surface code, we provide detailed methods of characterizing logical gate channels, 
including two-qubit gates implemented either transversally or via lattice surgery.
We also provide two protocols, exact and approximate, based on tensor network contraction~\cite{bravyi2014TND, chubb2021generaltensornetworkdecoding,kubica2024tensor} and minimum weight perfect matching (MWPM) 
decoding~\cite{Dennis2002topological, higgott2021pymatchingpythonpackagedecoding, wu2023fusionblossomfastmwpm}, respectively.
Furthermore, we propose three downstream applications of our characterization protocol 
in postselection, PEC, and ZNE.  Finally, through 
a numerical validation and resource estimation analysis, we show that our QEC+QEM architecture provides up to $87.4\%$ spacetime overhead 
savings compared to a plain QEC architecture and $65\%$ relative to GST. Our results set the foundation for the logical-level QEM workflow in the era of FTQC.
A qualitative comparison of our approach with other baseline methods is shown in Figure \ref{fig:overview}.

\textbf{Main contributions of our paper}
\begin{itemize}
\item We present an end-to-end protocol for logical gate error characterization using decoder soft information. Our estimator is unbiased, Rao-Blackwellized, and \emph{in situ} for any logical circuit. While our framework is general, we provide detailed characterization methods for surface code memory, single- and two-qubit gates (either transversal or via lattice surgery) to enable practical characterization of universal computation.
\item We introduce a soft-information-based post-selection and a runtime abort policy. The former reduces logical error rates by $100\times$ while discarding less than $0.1\%$ of the shots; the latter saves $50\%$ of the spacetime overhead on average.
\item We propose an efficient soft-information-based PEC pipeline for logical circuits. Our contributions include: (1) an adaptive shot allocation strategy that achieves optimal PEC outcome with limited resources, and (2) PEC on post-selected logical circuits to achieve significant shot overhead savings.
\item 
Using soft information, we efficiently characterize the logical channel for ZNE pipelines, demonstrating with an example of the noise amplification subroutine. 
\end{itemize}

The rest of the paper is structured as follows. In Section \ref{Background}, we introduce background on key theory and terms used in our paper. Section \ref{sec:estimation} presents the soft-information-based characterization protocol. We then introduce three main applications, including Section \ref{sec:post-selection}: post-selection and runtime abort policy, Section \ref{sec:PEC}: efficient logical PEC, and Section \ref{sec:ZNE}: efficient logical ZNE. Finally, we conduct comprehensive evaluations in Section \ref{sec:evaluations}, discuss related works in Section \ref{sec:related_works}, and conclude in Section \ref{sec:conlusion}.

\section{Background}
\label{Background}
\subsection{Quantum Error Correction}
As the fundamental building block of FTQC, a stabilizer code $\mathcal{S}$ encodes $k$ logical qubits into 
$n$ physical qubits with distance $d$, denoted by $[[n,k,d]]$ \cite{Shor1995decoherence, Steane1996, Gottesman1998FTtheory}. 
Through repeated syndrome extraction cycles, error information from the data qubits is projected onto ancilla qubits, enabling error detection 
and correction without revealing the encoded logical information. A classical decoding algorithm then maps the observed syndrome to an estimated error pattern 
for recovery. Such codes can detect up to $d-1$ errors and correct up to $\lfloor (d-1)/2 \rfloor$ arbitrary unknown errors. Crucially, the threshold theorem 
\cite{Aharonov1997,knill1996thresholdaccuracyquantumcomputation,Aliferis2006} guarantees that when physical error rates remain below a code-dependent threshold, 
logical error rates decrease exponentially with distance $d$, enabling scalable FTQC. 

\subsection{Surface Code}
The surface code is a planar, local stabilizer code with $X$- and $Z$-stabilizer checks  of weight at most four on a $2\text{D}$ lattice with open 
(rough/smooth) boundaries \cite{Kitaev2003FTQC, Dennis2002topological, Fowler2012surfacecode}.
Its high threshold and nearest-neighbor connectivity make it particularly practical for current hardware platforms, such as superconducting qubits.
A distance-$d$ surface code patch encodes one logical qubit with Hamming weight $d$ between complementary boundaries, and the logical operators $\bar{X}$ and $\bar{Z}$ correspond to chains of physical operators spanning these boundaries. The surface code naturally supports 
transversal Pauli operations and uses lattice surgery for two-qubit entangling gates via joint logical measurements, making the code
particularly suitable for universal quantum computation. 

The minimum weight perfect matching (MWPM) algorithm \cite{Dennis2002topological} is one of the most performant and widely 
used decoders for surface codes.
MWPM treats syndrome defects as vertices in a graph and finds the minimum-weight matching that pairs them, corresponding to the most likely error chain. For circuit-level noise models, MWPM extends naturally to 3D space-time decoding by matching syndrome defects across both spatial and temporal dimensions.
However, while MWPM is fast~\cite{higgott2021pymatchingpythonpackagedecoding,wu2023fusionblossomfastmwpm} and finds the most likely error chain, it ignores the physical error degeneracy and is therefore not a maximum likelihood (ML) decoder (unlike, for instance, the tensor network decoder~\cite{bravyi2014TND, chubb2021generaltensornetworkdecoding,kubica2024tensor}). 

\subsection{Quantum Error Mitigation}
QEM comprises a family of protocols that aim to reduce bias in observable expectation values with the knowledge of the noise channel. 
QEM has enabled practical quantum algorithms in the noisy intermediate-scale quantum era, such as the experimental demonstration of quantum utility \cite{Kim2023utility}. Unlike QEC, which tries to correct the quantum state itself, QEM seeks to average out the bias in the measurement statistics. Two main QEM protocols that are commonly used 
and will be discussed in this paper are PEC \cite{Temme2017QEM, PEC_lindbladian2023Ewout, PECscaling2021Andrea, PECdynamic2024Gupta, SimPEC2023Guimaraes} and ZNE \cite{Temme2017QEM, Giurgica_Tiron_2020ZNE, He2020ZNEinsert, Majumdar2023bestQEM}. 
PEC learns the noise channel $\mathcal{N}$ for each gate and implements its inverse $\mathcal{N}^{-1}$ via a signed quasi-probability decomposition. 
This requires sampling from a distribution with negative weights, leading to sampling complexity that scales exponentially with the circuit depth and noise strength. 
ZNE first amplifies noise by factors $\theta > 1$ and then extrapolates to $\theta = 0$ using Richardson extrapolation or polynomial fitting. 
While ZNE has polynomial overhead, it assumes smooth noise scaling and may fail for non-Markovian or correlated errors. Both 
protocols require accurate noise characterization and large shot overheads, become infeasible for large-scale applications, and are not fault-tolerant.

\subsection{Noise Characterization Protocol}
Accurate noise characterization is critical for both QEM protocols and quantum system calibration. Traditional approaches include 
gate-set tomography (GST) \cite{Nielsen_2021GST, Brieger2023CGST}, which reconstructs the complete process matrix for a gate set 
through self-consistent measurements, and quantum process tomography \cite{Huang_2020Classicalshadow}, which characterizes 
individual quantum channels. Both methods require exponential resources in system size and suffer from drift sensitivity 
\cite{wang2024drift, Proctor2020drift}. Randomized benchmarking \cite{Knill_2008RB, Hines2023RBdemo} and cycle benchmarking \cite{Zhang2025cb} provide more scalable alternatives but only estimate average gate fidelities rather than complete channel 
descriptions. 

While characterization is essential for QEM, existing protocols either incur large overhead for large-scale and fault-tolerant circuits
or lack robustness in handling noise drift. This motivates our approach of leveraging decoder soft information from 
QEC cycles to provide \emph{in situ} channel estimates without additional characterization experiments.
\section{Logical Error Estimation with Decoder Soft Information}
\label{sec:estimation}
\subsection{What is Decoder Soft Information?}
Traditional QEC studies treat syndrome decoding as a hard decision problem: given an observed syndrome $s$, 
the decoder outputs a single most-likely error pattern and applies the corresponding recovery operation. However, this 
approach discards valuable probabilistic information that could be leveraged for noise characterization and post-selection. 

Decoder soft information refers to this probabilistic output of a decoder that provides not just the most likely error 
pattern, but the complete probability distribution over all possible logical error outcomes given the observed syndrome. This can also be understood as the likelihood 
of a logical decoding failure or the confidence of the decoded result. For example, besides outputting a hard decision like ``an $X$ error on qubit $3$ occurred'', soft information provides probabilities that, based on this decoded result, the likelihood of logical decoding error  
``$P$(logical $X$) = $0.10$, $P$(logical $Y$) = $0.06$, $P$(logical $Z$) = $0.12$, $P$(no error) = $0.72$''.

To formalize this, let $\mathsf{synd}$ be the syndrome measurement map that takes any physical Pauli error and returns its syndrome outcome.
Let $\mathsf{corr}$ be the correction map that takes any valid syndrome and returns  deterministic but otherwise arbitrarily chosen Pauli error with that syndrome.
The operator $\bar{E} = E\cdot\mathsf{corr}(\mathsf{synd}(E))$ is a Pauli logical operator on the encoded logical qubit, i.e., $\bar{E} \in \{\bar I, \bar X, \bar Y, \bar Z\}$; we then also say that $E$ has a logical effect $\bar E$.
We partition all physical errors with a syndrome $s$ into four logical equivalence classes based on their logical effect.
Each class $\mathcal{C}_{s,\bar{L}}$ contains all physical error patterns that (1) produce syndrome $s$ and (2) have the same logical effect $\bar{L}$
\begin{equation}
\mathcal{C}_{s,\bar{L}} = \{E \in \mathcal{P}_n : \mathsf{synd}(E) = s \text{ and } \bar{E} = \bar{L}\}.
\label{eq:cosets}
\end{equation}
where $\mathcal{P}_n$ refers to the $n$-qubit Pauli group. 
Given a noise model that assigns probability $p(E)$ to each physical error pattern, the decoder computes the total probability 
of each logical class by summing over all physical errors in that class. The decoder soft output is then the normalized posterior probability over logical error classes
\begin{equation}
    p(\bar{L}\mid s)=\frac{\sum_{E\in\mathcal{C}_{s,\bar{L}}}p(E)}{\sum_{\bar{L}'\in\{I,X,Y,Z\}}\sum_{E\in\mathcal{C}_{s,\bar{L}'}}p(E)}.
\label{eq:posteriors}
\end{equation}
This posterior probability is the decoder soft information. By averaging these posteriors over many syndrome measurements obtained from actual circuit executions, we obtain an unbiased estimate of the underlying logical noise channel without requiring additional quantum experiments. The syndrome already contains all the information needed to characterize logical noise---we simply need to extract it properly, rather than discarding it through hard-decision decoding. Decoder soft information has been explored in QEC for applications such as concatenated code decoding, lattice surgery, and more~\cite{toshio2025decswitching, gidney2023yokedsurfacecodes, akahoshi2025runtimereductionLS, meister2024efficientsoftoutputdecoderssurface, Poulin_2006Optimaldecoding, Duclos_Cianci_2010}.

\subsection{Soft Information in Classical Computing}
The concept of soft information---probabilistic output instead of hard binary decisions---has been widely studied in classical computing. 
In classical communication settings, soft-decision decoding for error-correcting codes like LDPC and turbo codes significantly 
outperforms hard-decision decoding by propagating likelihood ratios through iterative belief propagation, providing confidence 
metrics for each decoded bit \cite{965575}. Similarly, in Bayesian machine learning and pattern recognition, classifiers output posterior 
probabilities $p(y|x)$ rather than just the most-likely class label, enabling uncertainty quantification and decision-making 
under risk. Medical diagnosis AI systems, for example, use these posteriors to assess confidence and combine multiple evidence 
sources \cite{Kompa2021SecondOpinion}.

\subsection{FTQC Model}
\begin{figure}[b]
    \centering
    \includegraphics[width=0.95\linewidth]{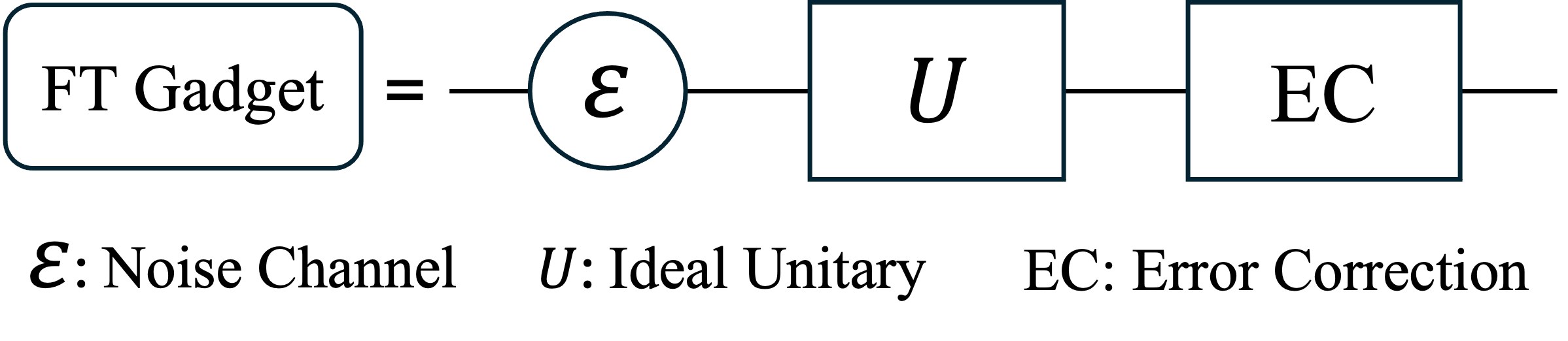}
    \caption{A fault-tolerant gadget for a unitary $U$ that we consider in this paper.}
    \label{fig:general_model}
\end{figure}

In this paper, we consider a simplified model of FTQC---each FT gadget is composed of a noise channel before the ideal unitary gate, followed by a QEC cycle, as depicted in Figure \ref{fig:general_model}.
The QEC cycle includes syndrome extraction (with no measurement errors), decoding, and recovery operations. 
In our studies, we assume that physical qubits are affected by a local stochastic noise channel, such as Pauli noise, which is also the standard for QEC studies~\cite{gottesman2009introductionquantumerrorcorrection, Suzuki2022qecqem}.
\subsection{Estimation Algorithms}
The core computational challenge in extracting decoder soft information is computing the posterior probability $p(\bar{L}|s)$ 
for each logical equivalence class $\bar{L} \in \{\bar I, \bar X, \bar Y, \bar Z\}$ given syndrome $s$. Recall from Equation \ref{eq:posteriors} that this requires 
calculating the total probability mass of all physical error patterns in each logical class, where the exact computation is a weighted counting problem over exponentially many error configurations. We present two 
approaches: an exact method using tensor network contraction and a fast approximate method using boundary-enforced MWPM.

\textbf{Exact Estimation via Tensor Network Contraction.} 
To compute exact posterior probabilities, we construct a two-dimensional tensor network that encodes both syndrome constraints 
and local error probabilities. Following the ML tensor network decoder framework, we represent the syndrome 
constraints using Kronecker-delta tensors $\delta$ that project onto error configurations consistent with the measured syndrome, 
and represent local Pauli error distributions using diagonal tensors $T_i$ for each data qubit $i$ with entries corresponding to 
$p(I_i), p(X_i), p(Y_i), p(Z_i)$. The partition function $Z_{s,\bar{L}}$ is obtained by contracting this tensor network
\begin{equation}
Z_{s,\bar{L}} = \text{Contract}\bigl(\{\delta\}\cup\{T_i\}_{i=1}^{n}\bigr)
\end{equation}

For a planar surface code lattice where $X$ and $Z$ errors are non-independent (e.g., depolarizing channel), exact contraction is \#P-complete with runtime $O(\exp(\Theta(\sqrt{n})))$ where $n = d^2$ is 
the number of physical qubits \cite{Schuch2007PEPS}. To improve efficiency, we adopt a matrix product state (MPS) approximation that sweeps through the 
lattice, absorbing tensors sequentially while truncating bond dimensions to $\chi$. This MPS approach 
achieves runtime $O(n\chi^3)$ while maintaining near-optimal accuracy with moderate $\chi$ 
(typically 16-32). The advantage of tensor network decoding is its exactness (or a controlled approximation with MPS). However, the polynomial scaling in bond dimension, exponential worst-case complexity, and limitation to code-capacity noise model
make it non-ideal for real-time decoding or large code distances \cite{bravyi2014TND, MarkovShi2008}.

\begin{figure}[t]
    \centering
    \includegraphics[width=\linewidth]{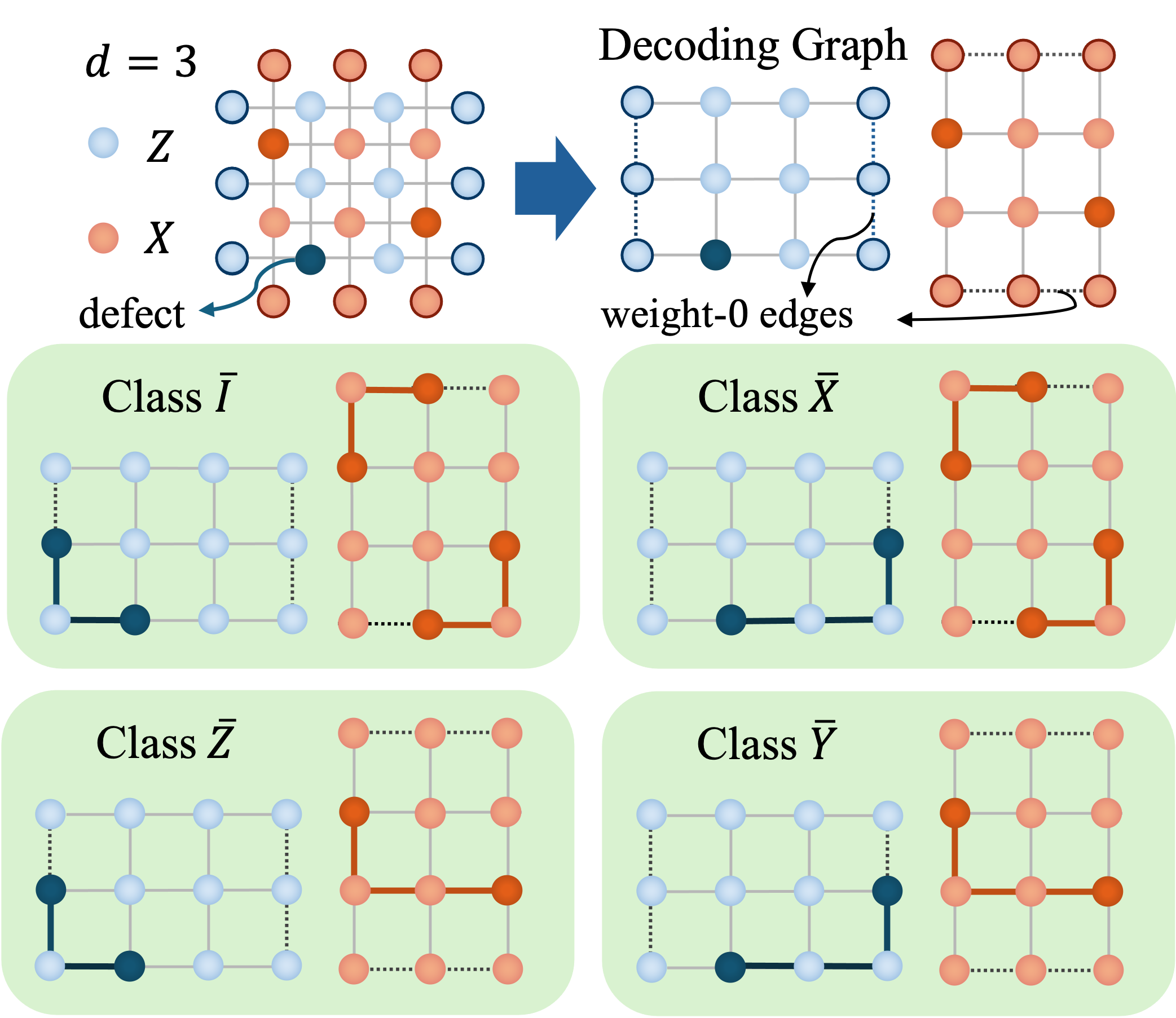}
    \caption{
    Boundary condition enforcement for a distance-3 surface code. Top: Decoding with virtual boundaries connected by zero-weight edges. Bottom: Four decoding graphs with different boundary 
    vertex types (defect/normal) corresponding to logical equivalence classes $\bar I$, $\bar X$, $\bar Z$, and $\bar Y$. Solid edges represent the 
    MWPM solution for each class.}
    \label{fig:boundary_enforce}
\end{figure}

\textbf{Fast Approximation via Boundary-Enforced MWPM.}
For practical deployment, we propose a fast approximate method based on MWPM with boundary condition 
enforcement. Logical operators for surface codes correspond to equivalence classes connecting complementary 
boundaries. By modifying the decoding graph to enforce specific boundary conditions, we can enforce MWPM to find the most-likely 
error within each logical equivalence class. Specifically, we consider an original decoding graph that consists of three types of vertices:
defect vertices, normal vertices, and virtual boundary vertices. MWPM algorithms will find the minimum weight solution that matches defect vertices an odd number of times, 
normal vertices an even number of times, and virtual boundary vertices an arbitrary number of times.  To enforce a specific logical class,
 we modify the graph construction as follows:

\textit{Graph Modification:} First, we connect all virtual vertices belonging to the same boundary type (e.g., all top and bottom 
$X$-boundary vertices) with zero-weight edges This is illustrated in 
Figure~\ref{fig:boundary_enforce}, where the modified graph shows virtual boundary vertices connected by dotted zero-weight edges. 
Second, we change the vertex type (defect vs. normal) of these merged boundary vertices to enforce the desired logical class. 
We can denote the default solution logical class as $\bar I$. To obtain posterior probabilities for all four 
logical equivalence classes $\{\bar I, \bar X, \bar Y, \bar Z\}$, we run MWPM with different boundary enforcement patterns and parity. For example, in Figure \ref{fig:boundary_enforce}, to enforce 
class $\bar X$, we set the $X$-boundaries from $(1,0)$ to $(0,1)$, $Z$-boundaries unchanged, and run MWPM. To enforce class $\bar Y$, we set both $X$- and $Z$-boundaries to defects with opposite parities to class $\bar I$.

Algorithm~\ref{alg:boundary_enforce} presents the pseudocode for boundary-enforced MWPM and the calculation of posterior probabilities. 
The algorithm takes as input 
the syndrome $s$, physical error rates $\{p_e\}$, and target logical class $\bar{L}$, and outputs the approximate posterior 
probability $\hat{p}(\bar{L}|s)$.

\begin{algorithm}[t]
\caption{Boundary-Enforced MWPM for Logical Posterior Estimation}
\label{alg:boundary_enforce}
\KwIn{Syndrome $s$, error rates $\{p_e\}$, logical class $\bar{L} \in \{I,X,Y,Z\}$}
\KwOut{Approximate posterior $\hat{p}(\bar{L}\mid s)$}
\tcp{Run default MWPM to find class $I$ and extract boundary parities}
$G \gets$ \textsc{BuildDecodingGraph}$(s, \{p_e\})$\;
$M \gets$ \textsc{MWPM}$(G)$\;
$(p_{X_1}, p_{X_2}) \gets$ \textsc{ExtractXBoundaryParities}$(M)$\;
$(p_{Z_1}, p_{Z_2}) \gets$ \textsc{ExtractZBoundaryParities}$(M)$\;
\If{$\bar{L} \neq I$}{
    $G \gets$ \textsc{ConnectBoundaryVertices}$(G)$\tcp*{Add zero-weight edges}
    $f_X \gets \mathbf{1}[\bar{L} \in \{X, Y\}]$, $f_Z \gets \mathbf{1}[\bar{L} \in \{Z, Y\}]$\;
    \textsc{SetBoundaryParities}$(G, X\text{-bdry}, (p_{X_1} \oplus f_X, p_{X_2} \oplus f_X))$\;
    \textsc{SetBoundaryParities}$(G, Z\text{-bdry}, (p_{Z_1} \oplus f_Z, p_{Z_2} \oplus f_Z))$\;
    $M \gets$ \textsc{MWPM}$(G)$\tcp*{Solve on modified graph}
}
$E_{\bar{L}} \gets$ \textsc{ExtractErrorPattern}$(M)$\;
$p_{\bar{L}} \gets \prod_{e \in E_{\bar{L}}} p_e \cdot \prod_{e' \notin E_{\bar{L}}} (1 - p_{e'})$\;
\Return{$p_{\bar{L}}$}
\end{algorithm}

To obtain all four posterior probabilities, we run Algorithm~\ref{alg:boundary_enforce} four times with different boundary 
configurations. The final normalized posteriors are:
$
\hat{p}(\bar{L}|s) = \frac{p_{\bar{L}}}{\sum_{\bar{L}' \in \{\bar I,\bar X,\bar Y,\bar Z\}} p_{\bar{L}'}}
$, where each $p_{\bar{L}}$ is the probability of the most-likely error pattern in that class, serving as an approximation to the 
full probabilities sum. We numerically study the bias from this approximation in section \ref{sec:evaluations}.

The runtime complexity for calculating posterior probabilities with MWPM is determined by the worst-case complexity of Edmonds' blossom algorithm: $O(d^6)$ for 2D syndrome graphs and $O(d^9)$ for 3D graphs \cite{wu2023fusionblossomfastmwpm, higgott2021pymatchingpythonpackagedecoding, Gabow1976, Lawler1976}. 

\subsection{Logical Memory and Transversal Single-Qubit Gates}
For logical memory operations and transversal single-qubit Pauli gates (natively supported by surface codes), the decoder soft 
information protocol directly applies Algorithm~\ref{alg:boundary_enforce} for a code-capacity noise model. In fault-tolerant circuits, syndrome extraction typically runs for $O(d)$ rounds to suppress measurement errors. For such multi-round scenarios, the decoding graph extends to three dimensions, with temporal edges connecting 
syndrome defects across rounds. The boundary enforcement strategy remains the same: we modify the 3D graph by connecting spatial 
boundary vertices and setting their types according to the target logical class \cite{gidney2023yokedsurfacecodes}. 

\subsection{Transversal Two-qubit Gates}
Decoding transversal two-qubit gates, such as the logical CNOT between two surface code patches, introduces additional complexity due to error 
propagation \cite{Beverland2021,sahay2025qeccnot, cain2025correlateddec, cain2025fastcorrelatedalgo}. Consider two distance-$d$ patches $q1$ (control) and $q2$ (target) undergoing a transversal 
CNOT, where physical CNOT gates are applied between corresponding data qubits. 

\begin{figure}[!t]
    \centering
    \includegraphics[width=\linewidth]{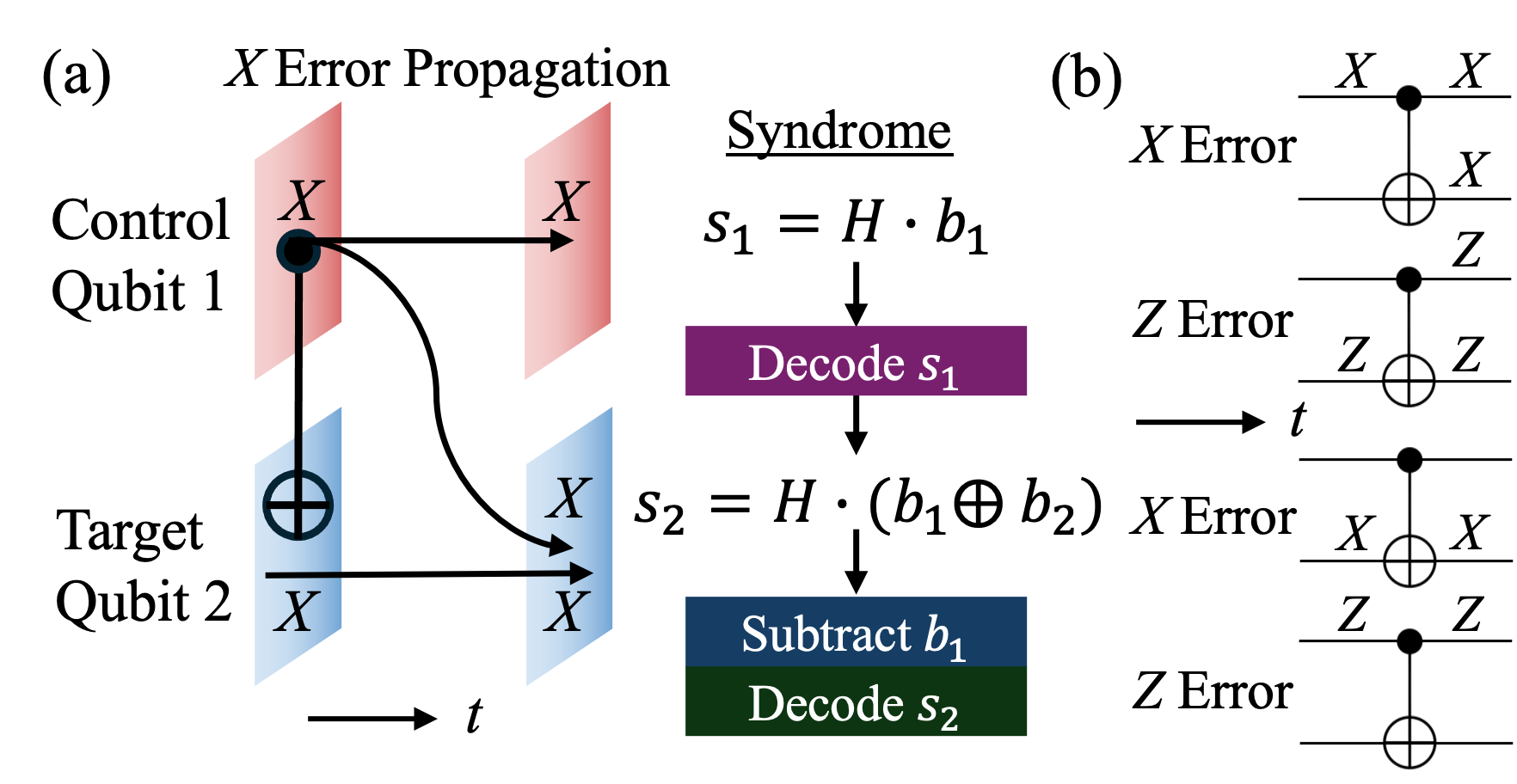}
    \caption{Transversal CNOT between two surface code patches of the same size. (a) Physical $X$ errors propagation from control to target qubits and sequential decoding strategy. (b) Error propagation for both $X$ and $Z$ errors on control and target qubits.} 
    \label{fig:tcnot}
\end{figure}

If physical errors occur independently after the transversal CNOT, the syndromes on each 
patch remain uncorrelated. In this case, we can apply Algorithm~\ref{alg:boundary_enforce} independently to each patch, obtaining 
two sets of posterior probabilities $\{p_{\bar{L}_1}\}$ and $\{p_{\bar{L}_2}\}$. The joint logical error probability for the 
two-qubit gate is simply the product: $p(\bar{L}_1 \otimes \bar{L}_2) = p(\bar{L}_1) \cdot p(\bar{L}_2)$. This yields 
$4 \times 4 = 16$ probabilities corresponding to all two-qubit Pauli operators.

When independent errors happen before or both before and after the 
transversal CNOT, error propagation through physical CNOTs creates syndrome correlations between the two patches. Specifically, 
$X$ errors on the control propagate as $X_1 \rightarrow X_1 X_2$, and $Z$ errors on the target propagate as 
$Z_2 \rightarrow Z_1 Z_2$. This produces coupled syndromes where the observed syndrome on one patch depends on errors from both 
patches shown in Figure \ref{fig:tcnot}.

To handle correlated errors, we employ \textit{sequential syndrome transfer decoding} \cite{cain2025correlateddec}. The key insight is that certain syndrome 
components remain independent despite propagation. For a CSS surface code, the syndrome equations become
\begin{align}
\mathbf{s}_{1,X} &= H_{1,X}\mathbf{b}_1, \quad \mathbf{s}_{2,Z} = H_{2,Z}\mathbf{a}_2 \\
\mathbf{s}_{1,Z} &= H_{1,Z}(\mathbf{a}_1 \oplus \mathbf{a}_2), \quad \mathbf{s}_{2,X} = H_{2,X}(\mathbf{b}_1 \oplus \mathbf{b}_2)
\end{align}
where $\mathbf{a}_k, \mathbf{b}_k$ are the $X$/$Z$ error bit-vectors on patch $k$. The sequential decoding procedure is:
\begin{enumerate}
    \item Decode $\mathbf{s}_{1,X}$ to obtain $\hat{\mathbf{b}}_1$ (control $X$-errors)
    \item Decode $\mathbf{s}_{2,Z}$ to obtain $\hat{\mathbf{a}}_2$ (target $Z$-errors)
    \item Compute residual syndromes by subtracting propagated contributions:
    $\mathbf{s}_{1,Z}^{\text{res}} = \mathbf{s}_{1,Z} \oplus H_{1,Z}\hat{\mathbf{a}}_2$ and 
    $\mathbf{s}_{2,X}^{\text{res}} = \mathbf{s}_{2,X} \oplus H_{2,X}\hat{\mathbf{b}}_1$.
    \item Decode residual syndromes to obtain $\hat{\mathbf{a}}_1$ and $\hat{\mathbf{b}}_2$
\end{enumerate}

For soft information extraction with correlated errors, we apply boundary-enforced MWPM at each decoding step. For example, when 
decoding $\mathbf{s}_{1,X}$ for control $X$-errors, we run Algorithm~\ref{alg:boundary_enforce} four times with different 
$X$-boundary conditions to obtain $p(\bar{L}_{1,X})$. Similarly, we decode $\mathbf{s}_{2,Z}$ to get $p(\bar{L}_{2,Z})$. After 
syndrome transfer, we decode the residual syndromes to obtain $p(\bar{L}_{1,Z})$ and $p(\bar{L}_{2,X})$. The joint two-qubit 
logical error distribution is then approximated by combining these marginal distributions, accounting for the correlation structure 
imposed by error propagation.

\subsection{Lattice Surgery}
Lattice surgery \cite{Horsman_2012LSsurface, fowler2019LSlow-overhead, Erhard_2021LSentangled, Chamberland2022TELS, Litinski2019gamesurfaceLS} provides an alternative approach to implementing two-qubit logical gates that requires only nearest-neighbor 
interactions, making it more hardware-friendly than transversal gates. Instead of directly moving qubits from different patches, lattice surgery performs joint measurements between adjacent boundaries of code patches, effectively ``merging'' and ``splitting'' the 
logical qubits. In this section, we use a lattice surgery-implemented CNOT gate as an example to demonstrate the soft information calculation.

\begin{figure}[!t]
    \centering
    \includegraphics[trim={0 0.5cm 0 0}, width=\linewidth]{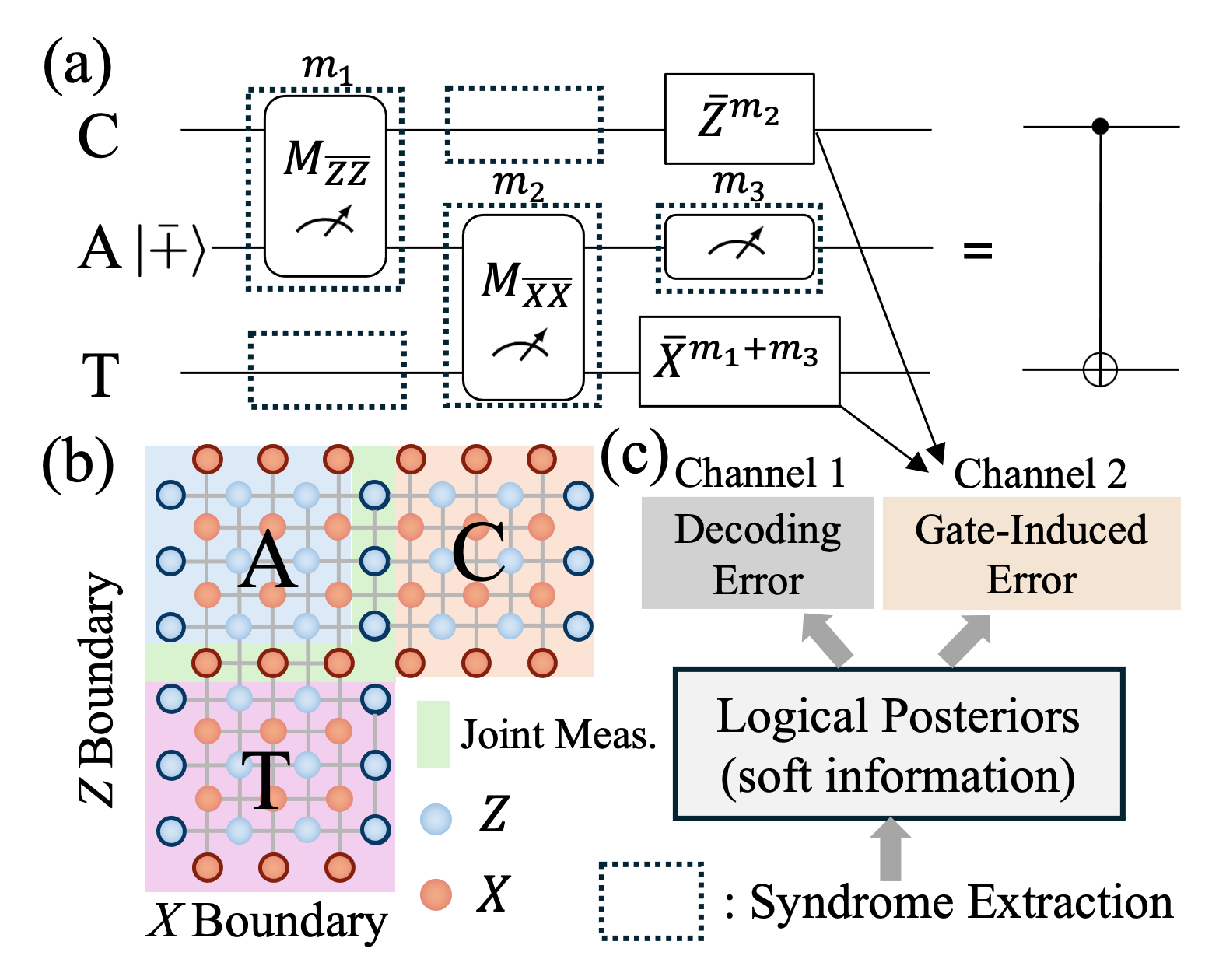}
    \caption{
    Fault-tolerant logical CNOT via lattice surgery.
    (a) Circuit diagram and syndrome extraction cycle (dashed box).
    (b) Physical layout of the three patches involved.
    (c) Breakdown of the logical error channel.}
    \label{fig:lattice_surgery}
\end{figure}

\textbf{Fault-Tolerant Lattice Surgery CNOT.}
To implement a logical CNOT gate between control qubit $\ket{\overline{c}}$ and target qubit $\ket{\overline{t}}$ using lattice surgery, we introduce 
an intermediate ancilla patch initialized in the $\ket{\overline{+}}$ state, as shown in Figure~\ref{fig:lattice_surgery}. Then, three joint logical measurements are performed:
\begin{enumerate}
    \item Joint $\overline{ZZ}$ measurement between control $\ket{\overline{c}}$ and ancilla $\ket{\overline{+}}$, yielding measurement outcome $m_1$;
    \item Joint $\overline{XX}$ measurement between ancilla $\ket{\overline{+}}$ and target $\ket{\overline{t}}$, yielding measurement outcome $m_2$;
    \item $\overline{Z}$ measurement on the ancilla, yielding outcome $m_3$.
\end{enumerate}
Each joint measurement is performed by merging the appropriate boundaries of adjacent patches, measuring the appropriate stabilizers, and 
decoding the resulting syndromes. Depending on the measurement outcomes, Pauli corrections are applied: $\bar{Z}^{m_2}$ applied to the control qubits patch and $\bar{X}^{m_1+m_3}$ applied to the target qubit patch. 

\textbf{Decoding Error.}
In a lattice surgery CNOT, we require all steps to be fault-tolerant, hence syndrome extractions + decoding (dashed box in Figure \ref{fig:lattice_surgery}) are performed in each logical measurement and idle period. Consequently, each syndrome extraction will incur a decoding error channel characterized by decoder soft information. We denote the posterior probabilities from the $i$-th decoding step as $p_{i,a}$ where $i \in \{1,2,3\}$ 
indexes the measurement stage and $a \in \{\bar{X},\bar{Y},\bar{Z}\}$ indexes the logical Pauli error type.

For each decoding step, we apply Algorithm~\ref{alg:boundary_enforce} to the joint decoding graph (which includes both patches 
involved in that measurement). This yields the following logical error probabilities 
\begin{itemize}
    \item $p_{1,\bar X}, p_{1,\bar Y}, p_{1,\bar Z}$---from $M_{\bar{ZZ}}$;
    \item $p_{2,\bar X}, p_{2,\bar Y}, p_{2,\bar Z}$---from $M_{\bar{XX}}$;
    \item $p_{3,\bar X}, p_{3,\bar Y}, p_{3,\bar Z}$---from $M_{\bar{Z}}$.
\end{itemize}
Apart from these, one can also calculate the logical error probabilities associated with the two idle period syndrome extraction and decoding steps. These Pauli errors are the first part of error channels depicted in Figure \ref{fig:lattice_surgery}(c) caused by decoding errors in lattice surgery CNOT. 

\textbf{Gate-Induced Error.}
Besides the decoding error channels discussed above, the incorrect measurement results ($m_1, m_2$ and $m_3$) will lead to incorrect conditional operations ($\bar{Z}^{m_2}$ and $\bar{X}^{m_1+m_3}$) and induce a second part of the error channel (gate-induced error in Figure \ref{fig:lattice_surgery} (c)). Since this channel solely comes from incorrect Pauli-$\bar{Z}$ and Pauli-$\bar{X}$ gates, we can write the effective two-qubit gate error channel as
\begin{align}
&\mathcal{N}_{\text{gate-induced}}(\rho) = p_{\bar{II}}\rho + p_{\bar{ZI}}(\bar{Z}\otimes \bar{I})\rho(\bar{Z}\otimes \bar{I}) \nonumber 
+ \\ &p_{\bar{IX}}(\bar{I}\otimes \bar{X})\rho(\bar{I}\otimes \bar{X}) + p_{\bar{ZX}}(\bar{Z}\otimes \bar{X})\rho(\bar{Z}\otimes \bar{X})
\label{eq:gate-induced_error}
\end{align}
where $p_{\bar{II}} = 1 - p_{\bar{ZI}} - p_{\bar{IX}} - p_{\bar{ZX}}$. The error rate are computed from measurement error probabilities $p_{m_i}$ via
\begin{align}
p_{\bar{ZI}} &= p_{m_2}(1-p_{m_1})(1-p_{m_3}) + (1-p_{m_2})p_{m_1}p_{m_3}, \nonumber \\
p_{\bar{IX}} &= (1-p_{m_2})(1-p_{m_1})p_{m_3} + (1-p_{m_2})p_{m_1}(1-p_{m_3}), \nonumber \\
p_{\bar{ZX}} &= p_{m_2}(1-p_{m_1})p_{m_3} + p_{m_2}p_{m_1}(1-p_{m_3}).
\end{align}
The measurement error probabilities are derived from the soft information posteriors in the decoding step. For measurement $i$, the error probability 
is the sum of logical error probabilities that would flip the measurement outcome:
$ p_{m_1} = p_{1,\bar{X}} + p_{1,\bar{Y}},\, p_{m_2} = p_{2,\bar{Z}} + p_{2,\bar{Y}},\, p_{m_3} = p_{3,\bar{Z}} + p_{3,\bar{Y}}.
$ By substituting these measurement error probabilities into Formula \ref{eq:gate-induced_error}, we obtain the gate-induced error channel. Since both decoding and gate-induced channels are Pauli, we can write the final effective noise channel for the lattice surgery CNOT as a composite of the two by following commutativity. Formally, the entire 
channel can be expressed as $\mathcal{N}_{\text{total}} = \mathcal{N}_{\text{decoding}} \circ \mathcal{N}_{\text{gate-induced}}$ and both components are directly estimated from soft information, enabling 
complete \textit{in situ} characterization of the lattice surgery operation.
\section{Application 1: Post-selection and Runtime Abort Policy}
\label{sec:post-selection}
Post-selection is a widely used technique in quantum computing to improve output fidelity by discarding measurement outcomes that 
are likely erroneous. Traditional post-selection methods rely on simple metrics like detection events. However, 
decoder soft information provides a much more nuanced and agile approach to post-select error-corrected circuits.

\textbf{Post-Selection Protocol.}
Consider a fault-tolerant quantum circuit executed with $N$ shots. For each shot $i$ and each logical gate $g$ in the circuit, we 
compute the posterior probability distribution $\{p_{i,g}(\bar{L})\}_{\bar{L} \in \{\bar{I}, \bar{X}, \bar{Y}, \bar{Z}\}}$ using 
Algorithm~\ref{alg:boundary_enforce}. The probability of a logical error occurring at gate $g$ in shot $i$ is
$p_{i,g}^{\text{err}} = 1 - p_{i,g}(I) = p_{i,g}(X) + p_{i,g}(Y) + p_{i,g}(Z)$.
We define a confidence threshold $\tau$ and discard shot $i$ if $\max_{g} p_{i,g}^{\text{err}} > \tau$.
Alternatively, we can use the accumulated error probability across all gates in the circuit and discard shot $i$ if
$\sum_{g=1}^{N_G} p_{i,g}^{\text{err}} > \tau_{\text{total}}$, where $N_G$ is the total number of logical gates in the circuit.
The logical error rate improvement is thoroughly analyzed in Section~\ref{sec:evaluations}.

\textbf{Runtime Abort Policy.}
When soft information can be computed in real-time (with complexity linear to MWPM decoding), we can abort circuit execution mid-shot upon detecting high logical error probability. If the accumulated error probability $\sum_{g=1}^{g_{\text{current}}} p_{i,g}^{\text{err}}$ exceeds threshold $\tau_{\text{abort}}$ at any point during execution, we terminate the remaining gates for that shot. This saves quantum spacetime volume by discarding shot that is likely to fail early.
\section{Application 2: Efficient Logical PEC}
\label{sec:PEC}
Before diving into the logical PEC protocol, we first review the unbiased channel characterization capability via decoder soft information.

\textbf{Unbiased Estimator.} Assume syndromes are generated from i.i.d.\ physical noise across $N$ shots.
Then the posterior $p(\bar L=\ell\mid s)$, viewed as a random variable over the syndrome $s$, has
expectation equal to the true logical-class probability $p_\ell \equiv \Pr(\bar L=\ell)$.
This follows by marginalization
$\mathbb{E}_{s}\!\left[p(\bar L=\ell \mid s)\right]
= \sum_{s} p(s)\,p(\bar L=\ell\mid s)
= \sum_{s} p(\bar L=\ell, s)
= p_\ell .$
Therefore, the sample average
$\hat p_\ell = \frac{1}{N}\sum_{i=1}^N p(\bar L=\ell\mid s_i)$ is an unbiased estimator:
$\mathbb{E}[\hat p_\ell]=p_\ell$.

Moreover, this estimator is the Rao-Blackwellized
version of the naive empirical estimator 
$\frac{1}{N}\sum_{i=1}^N \mathbf{1}_{\bar{L},i}$, and thus has provably equal or lower variance than the empirical estimator (e.g., GST protocols):
$\text{Var}(\hat{p}_{\bar{L}}) \leq \frac{1}{N}p_{\bar{L}}(1-p_{\bar{L}})$. This builds the theoretical foundation that enables accurate and efficient channel characterization via arbitrary circuits without additional quantum characterization experiments.

\textbf{PEC on Logical Circuits.}
PEC represents the inverse of each noisy logical gate $\mathcal{N}$ as a quasi-probability mixture: $\mathcal{N}^{-1} = \sum_i \eta_i B_i$ where 
$\{B_i\}$ are implementable Pauli operations and $\{\eta_i\}$ are real coefficients (possibly negative) with $\sum_i \eta_i = 1$. 
While $\mathcal{N}^{-1}$ is not a physical quantum channel (due to negative coefficients), it can be 
implemented in expectation through Monte Carlo sampling. For each shot, we sample operation $B_i$ with probability 
$|\eta_i|/\gamma$ where $\gamma = \sum_i |\eta_i|$ is the sampling overhead, and record the sign $\text{sgn}(\eta_i)$. The final 
expectation value is computed as a signed average over all samples, yielding an unbiased estimator whose variance scales as 
$\gamma^{2N_G}$ for a circuit with $N_G$ gates.

In logical circuits, QEC suppresses physical errors, leaving residual logical errors from rare high-weight events that PEC can mitigate without requiring high-distance codes. Since PEC overhead scales exponentially with gate count and error rate, logical PEC is highly favorable: QEC achieves low error rates (e.g., $10^{-5}$) and we mitigate only logical gates rather than quadratically more physical gates. Our protocol uses soft information from algorithm circuits---not specialized characterization---to estimate logical error channels. Syndrome measurements during normal execution provide posterior probabilities that, averaged across shots, yield unbiased channel estimates reflecting actual algorithm noise, eliminating separate characterization and avoiding drift-related inaccuracies.

\begin{figure}[!t]
    \centering
    \includegraphics[width=\linewidth]{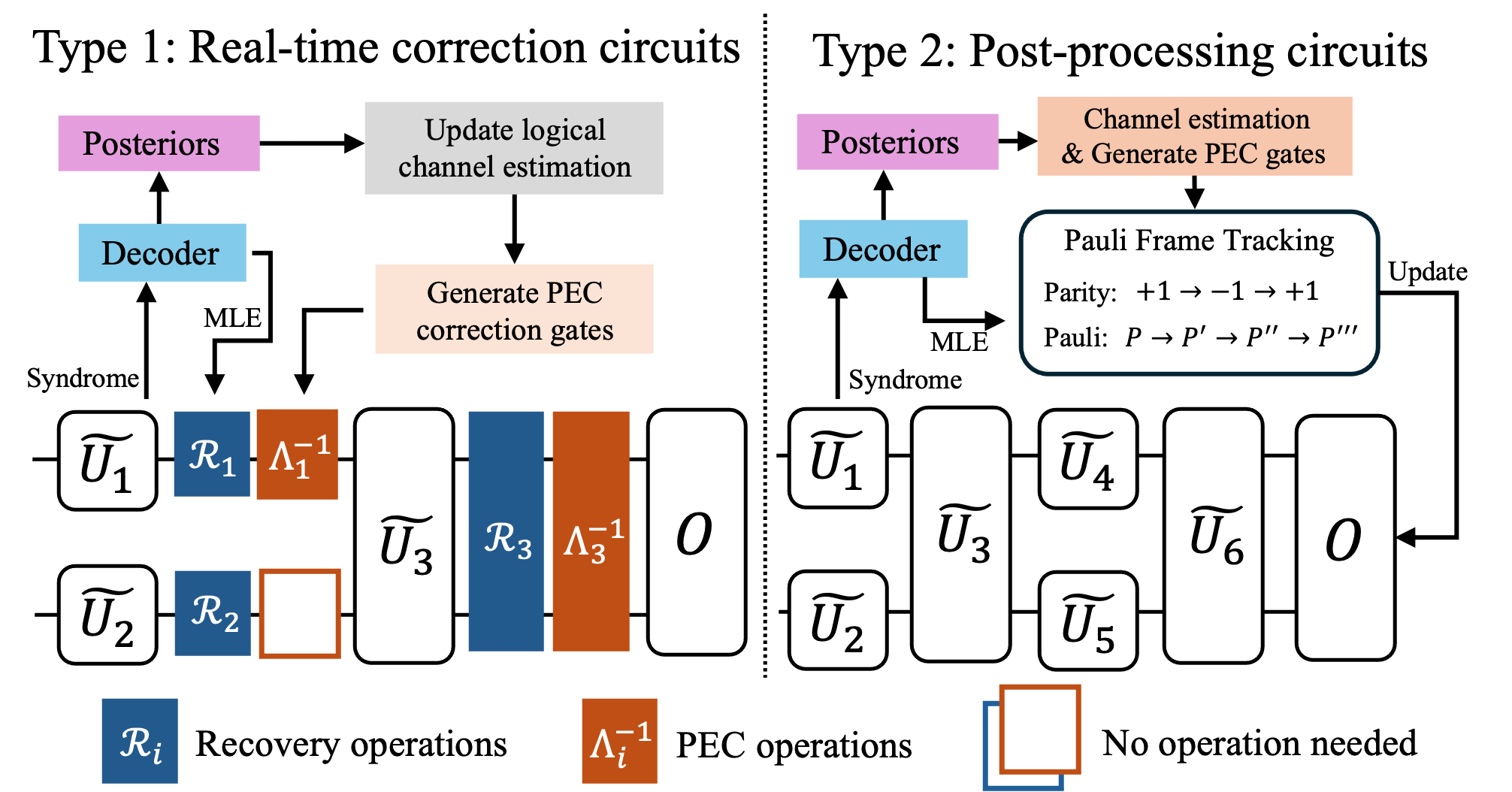}
    \caption{PEC implementation pipelines for two types of circuits. Type 1 (left): Real-time correction for circuits requiring immediate 
    feedback, with online channel estimation, recovery, and PEC gate insertion. Type 2 (right): Post-processing for Clifford circuits, using 
    Pauli frame tracking to apply corrections with software after all shots are complete.}
    \label{fig:pec_pipeline}
\end{figure}

\textbf{Optimal Shot Allocation Strategy.}
Given a fixed budget of $N$ shots, an important question is how to optimally allocate them between soft-information-based characterization and mitigation. 
Figure~\ref{fig:pec_pipeline} illustrates two implementation strategies depending on circuit requirements.

\textit{Type 1: Real-Time Correction Circuits.} For circuits requiring real-time decoding (e.g., those with $T$-gate injection via 
magic state distillation), we employ an adaptive strategy:
\begin{enumerate}
    \item Execute the algorithm for $k$ shots ($k \ll N$) to obtain initial channel estimates. These shots produce unmitigated but 
    still useful results.
    \item Once the estimated error rates stabilize (typically $k \approx 1000$-$10000$ shots), begin applying PEC corrections 
    starting from shot $k+1$.
    \item Continuously update channel estimates via an online running average: 
    $\hat{p}_{\bar{L}}^{(n)} = \frac{n-1}{n}\hat{p}_{\bar{L}}^{(n-1)} + \frac{1}{n}p(\bar{L}_n|s_n)$, refining the logical error rate estimation as more data accumulates.
\end{enumerate}

This adaptive approach balances characterization and mitigation: early shots characterize the noise while still contributing to the 
algorithmic output, and later shots benefit from increasingly accurate mitigation. The PEC correction gates (sampled Pauli operations $B_i$) need to be inserted into the circuit after 
each logical gate based on the current channel estimate and the sampled quasi-probability index. However, since the PEC gates are all Pauli gates,
the insertion can be done in software without additional quantum operations.

\textit{Type 2: Post-Processing Circuits.} For pure Clifford circuits or algorithms where post-processing suffices, we use a 
simpler approach:
\begin{enumerate}
    \item Execute the algorithm for all $N$ shots without mitigation.
    \item After completion, estimate logical error rates using soft information from all shots.
    \item Apply PEC corrections via Pauli frame tracking in classical post-processing, updating measurement outcomes according to 
    the quasi-probability sampling.
\end{enumerate}

This post-processing approach is particularly efficient for Clifford circuits because Pauli errors can be tracked classically and Pauli corrections can be applied without additional quantum operations. Given the estimated channel for each 
gate, we retroactively sample PEC correction operations and propagate them through the circuit using Pauli frame updates. The final 
measurement outcomes are adjusted based on the accumulated Pauli frame and the signs of the sampled quasi-probabilities. This 
purely classical post-processing incurs \textbf{zero quantum overhead} beyond the original $N$ shots, making it highly 
resource-efficient. The key advantage is that all $N$ shots contribute to both the channel estimation (minimum bias) and the 
final mitigated expectation value (minimum variance).

\begin{figure}[!t]
    \centering
    \includegraphics[trim={0 0.4cm 0 0}, width=\linewidth]{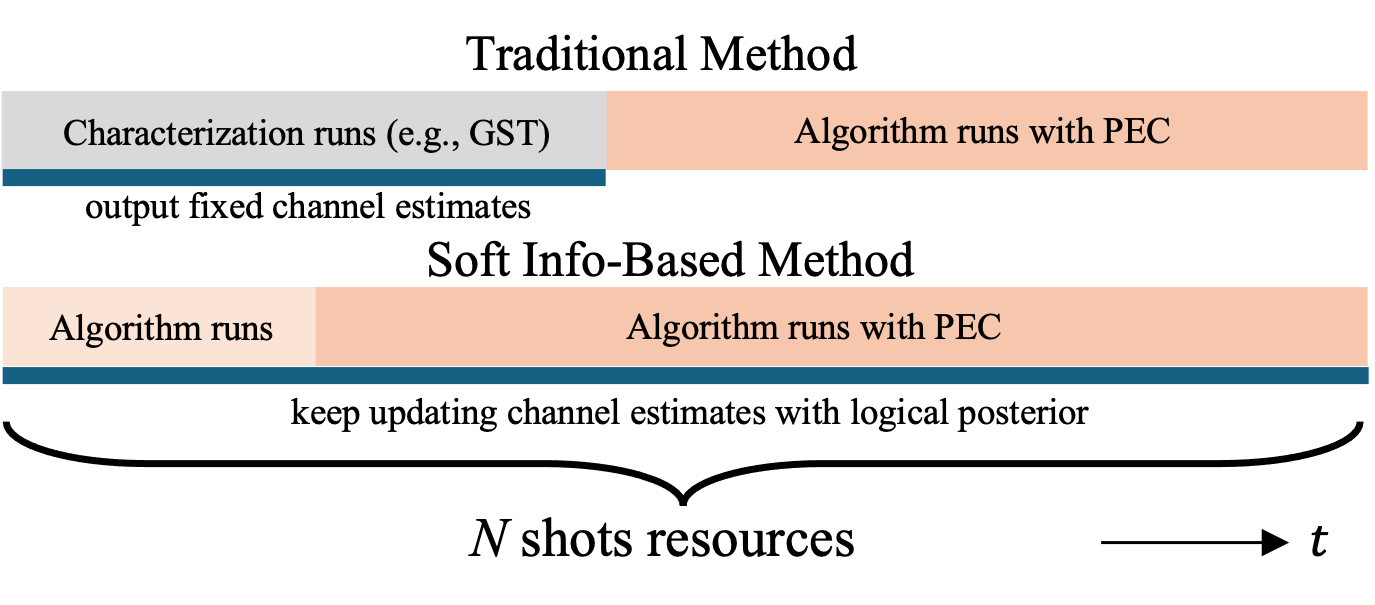}
    \caption{
    Visualization of the shot usage advantage of the soft-information-based method (ours) compared to the traditional method. With given resources for shots, our protocol does not require characterization circuit and can keep updating channel estimates with more algorithm experiments.}
    \label{fig:shotusage_pec}
\end{figure}

\textbf{Advantages Over Traditional PEC.}
Traditional PEC implementations using GST or other characterization protocols suffer from two major drawbacks. First, they require 
separate characterization and mitigation phases with no overlap, as shown in Figure~\ref{fig:shotusage_pec}. If $N_{\text{char}}$ 
shots are used for characterization, only $N - N_{\text{char}}$ shots remain for the actual algorithm. Second, noise drift between 
characterization and execution phases degrades mitigation accuracy.

Our soft-information-based approach eliminates both issues. All $N$ shots contribute to both algorithm execution and channel 
estimation, maximizing resource utilization. Channel estimates are continuously updated from the algorithm circuits themselves, 
ensuring they reflect current noise conditions and eliminating drift errors. This \textit{in-situ} characterization is particularly 
valuable for long-running algorithms or systems with time-varying noise.

\textbf{PEC on Post-Selected Circuits for Shot Overhead Saving.}
The combination of post-selection and PEC can achieve logical PEC efficiency beyond what either technique provides individually, 
leading to dramatic reductions in PEC sampling overhead. Since soft-information-informed post-selection reduces the logical error rates 
that PEC takes to invert, it also lowers the sampling overhead factor $\gamma$ exponentially.

The sampling overhead for PEC on a circuit with $N_G$ gates scales as \cite{Suzuki2022qecqem}:
\begin{equation}
\text{Overhead}_{\text{PEC}} \approx \exp(4N_G p_{\text{avg}})
\end{equation}
where $p_{\text{avg}}$ is the average logical error rate per gate. This exponential scaling makes PEC prohibitively expensive for 
circuits with high error rates or many gates. However, post-selection using soft information can reduce $p_{\text{avg}}$ by orders 
of magnitude with only a small fraction of rejected shots.

\emph{Example.} Consider a circuit with $N_G = 10{,}000$ gates and baseline logical error rate $p_L = 5 \times 10^{-5}$ per gate. Apply 
post-selection with discard rate $d = 0.01\%$, achieving $R = 10\times$ error rate improvement (numerically verified data point). The shot overhead factor 
comparing PEC with post-selection to PEC without post-selection is:
\begin{equation}
\eta = \frac{\exp\left[-4N_G p_L(1-1/R)\right]}{(1-d)^{N_G}} 
= \frac{e^{-1.8}}{e^{-1}} = e^{-0.8} \approx 0.45
\label{eq:pec_postselect}
\end{equation}

The 10× error rate improvement provides PEC savings ($e^{-1.8}$) that exceed the compounding discard overhead across 10,000 gates 
($e^{-1}$), resulting in $>50\%$ net resource savings. This benefit from soft information-based post-selection makes our logical PEC protocol even more resource-efficient. 
\section{Application 3: Efficient Logical ZNE}
\label{sec:ZNE}
ZNE suppresses errors by amplifying noise to scaling factors $\theta_j = s_j\theta$ and 
extrapolating back to the zero-noise limit \cite{He2020ZNEinsert,Temme2017QEM,Wahl2023ZNEscaling}. The noisy expectation admits expansion $\langle O \rangle_{\theta} = \sum_{k=0}^{\infty} a_k \theta^k$ where $a_0$ is the ideal value. Measuring at multiple noise levels and constructing a Richardson estimator 
$\hat{\langle O \rangle}_{\text{ZNE}} = \sum_{j=0}^{K} c_j \langle O \rangle_{\theta_j}$ with $\sum_{j=0}^{K} c_j s_j^k = \delta_{k0}$ 
cancels the first $K$ terms, achieving $O(\theta^{K+1})$ error and polynomial sampling overhead $\sum_j |c_j|$.
Traditional noise amplification via unitary folding ($U \to U(U^\dagger U)^m$) increases circuit depth and assumes uniform scaling. 
Probabilistic error amplification (PEA) offers a superior alternative using soft information channel estimates \cite{Kandala_2019QEMextends, Kim2023utility}. Given 
$\mathcal{N}_{\theta}(\rho) = (1-p_{\text{tot}})\rho + \sum_{P \neq I} p_P P\rho P$, we reconstruct the amplified channel
$\mathcal{N}_{s\theta} = (1-sp_{\text{tot}})\mathcal{I} + s \sum_{P \neq I} p_P \mathcal{P}$ by probabilistically inserting Pauli 
operations with scaled probabilities $\{sp_P\}$. This requires only single-qubit gates, accurately models 
gate-dependent noise, and adapts to time-varying channels through continuous soft information updates.

\begin{figure}[t]
    \centering
    \includegraphics[width=\linewidth]{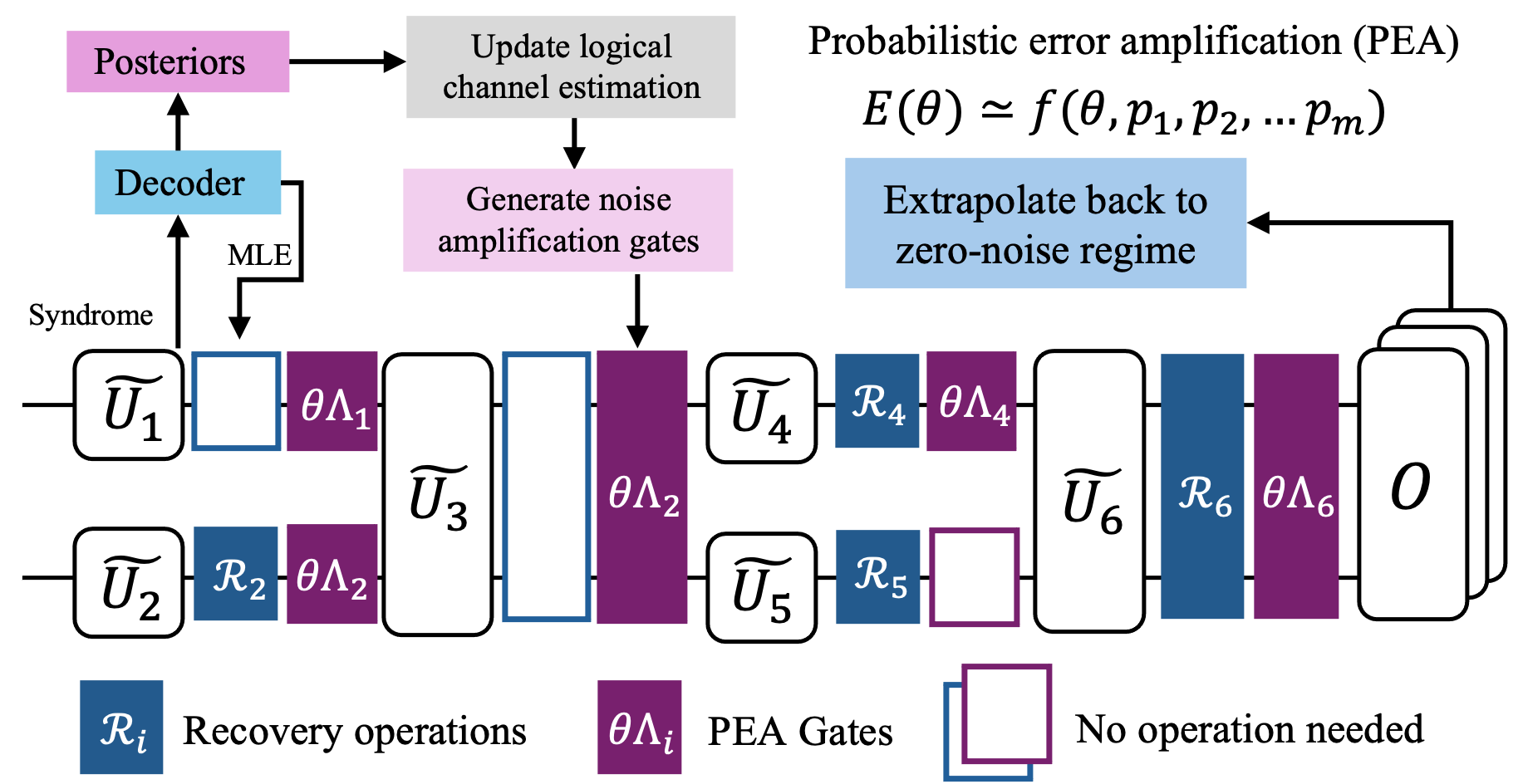}
    \caption{Logical ZNE pipeline with probabilistic error amplification. After each logical gate, the decoder computes soft 
    information (posteriors) to estimate the logical channel. These estimates are used to generate noise amplification gates 
    $\{\theta\Lambda_i\}$ that are inserted into the circuit. The observable is measured at multiple amplification factors $\theta_j$ 
    and extrapolated back to the zero-noise regime $\theta = 0$.}
    \label{fig:zne_pipeline}
\end{figure}

\textbf{Logical ZNE Pipeline with Soft Information.}
As illustrated in Figure~\ref{fig:zne_pipeline}, the pipeline integrates soft information extraction with noise amplification. 
During circuit execution with syndrome extraction, the decoder computes posterior probabilities $\{p(\bar{L}|s)\}$ after each 
logical gate. These posteriors are accumulated across shots to estimate the per-gate logical channel: 
$\hat{p}_{\bar{L},g} = \frac{1}{n}\sum_{i=1}^n p(\bar{L}_{i,g}|s_{i,g})$. For noise amplification, the estimated error rates are 
scaled by target factors $\theta_j$ to generate probabilistic error amplification (PEA) gates, which are inserted as Pauli 
operations after each logical gate. The observable is measured at multiple amplification factors $\{\theta_0 = 1, \theta_1, \ldots, \theta_K\}$, 
and Richardson extrapolation yields the zero-noise estimate $\hat{\langle O \rangle}_{\text{ZNE}}$. Similar to PEC, the key advantage is that channel estimation happens simultaneously with algorithm execution, eliminating separate 
characterization overhead and avoiding noise drift issues. 
\section{Evaluations}
\label{sec:evaluations}
\subsection{QEM Validation}
We first validate the efficacy of all three QEM applications introduced via numerical simulations, shown in Figure \ref{fig:post_selection}. 

\textbf{Logical PEC}: We simulated $100$ random Clifford circuits consisting of $n_q=100$ logical qubits encoded in $d=5$ surface codes with a physical error rate of $p=0.01$. Clifford circuits are generated with an ideal and deterministic observable expectation value of $\langle O \rangle_{\text{ideal}} = 1.0$ for a non-trivial Pauli string. We compare in Figure \ref{fig:post_selection}(a) the expectation distribution of unmitigated, PEC-MWPM, PEC-TN. 

\textbf{Logical ZNE.} For the logical ZNE protocol, shown in Figure \ref{fig:post_selection}(b), we scale the noise with a scaling factor $\theta\in\{1,2,...,9\}$ for distance $d\in\{3,5\}$ and compare this between ZNE-MWPM and ZNE-TN. 

\textbf{Post-selection.} We simulated surface code patches with distance $d\in\{ 3,5,7,9\}$ with physical error rate $p=0.01$ for $10^6$ shots. For each shot, we compute the logical $X$ posterior probability from the decoder soft information. Figure \ref{fig:post_selection}(c) shows the distribution plotted with $-\log(p_X)$. By discarding all samples $<2$, we discarded less than $0.01\%$ of samples, but the rest of the samples have logical error rates more than $9$ times smaller than the original samples. This effect is even more pronounced with higher distances and higher discard rates. In Figure \ref{fig:post_selection}(d), the posterior improvement ratio ($p_{\text{original}}/p_{\text{post-selected}}$), and equivalently logical error rate improvement ratio, improves up to $10^4$ with less than $1\%$ discard rate. We expect the logical error rate improvement to be even better with lower physical error rates.
\begin{figure}[!b]
    \centering
    \includegraphics[trim={0 0.7cm 0 0},width=\linewidth]{figures/QEM-ALL.png}
    \caption{Validation of QEM protocols. (a) Logical PEC results for a $d=5$ surface code at physical error rate $p=0.01$. (b) Logical ZNE results for $d\in\{3,5\}$ at $p=0.01$. (c) Soft information-based post-selection: histogram of log-scale logical-$X$ posteriors for $d=5$, $p=0.01$, showing $>9\times$ improvement with $<0.01\%$ of shots discarded (cutoff threshold $= 2$). (d) Posterior improvement ratio versus discard rate for various code distances at $p=0.01$.}
    \label{fig:post_selection}
\end{figure}

\textbf{Duration Savings from Runtime Abort.}
Consider a process of \(N\) steps where each step independently aborts with probability \(p\in(0,1)\). Let \(K\in\{1,\dots,N\}\) denote the first abort step (if any), and define saved steps as \(S:= N-K\).

The first abort occurs at step \(k\) with probability \(\Pr(K=k)=(1-p)^{k-1}p\). Conditioning on at least one abort occurring, we have
\begin{equation}
\Pr(K=k\mid \text{abort})=\frac{(1-p)^{k-1}p}{1-(1-p)^N},\quad k=1,\dots,N.
\end{equation}
Using the finite geometric series identity \(\sum_{k=1}^N k\,q^{k-1}=[1-(N+1)q^N+Nq^{N+1}]/(1-q)^2\) with \(q=1-p\), we obtain
\begin{equation}
\mathbb{E}[S\mid \text{abort}]
= N-\frac{1-(N+1)(1-p)^N+N(1-p)^{N+1}}{p\,[1-(1-p)^N]}
.
\end{equation}

\noindent\emph{Asymptotic behavior:} For rare aborts (\(pN\ll 1\)), \(\mathbb{E}[S\mid \text{abort}] \approx N/2\); for fixed \(p>0\) and large \(N\), \(\mathbb{E}[S\mid \text{abort}]\to N-1/p\).

\subsection{PEC Strategies}

\begin{figure}[!t]
    \centering
    \includegraphics[trim={0 0.5cm 0 0},width=\columnwidth]{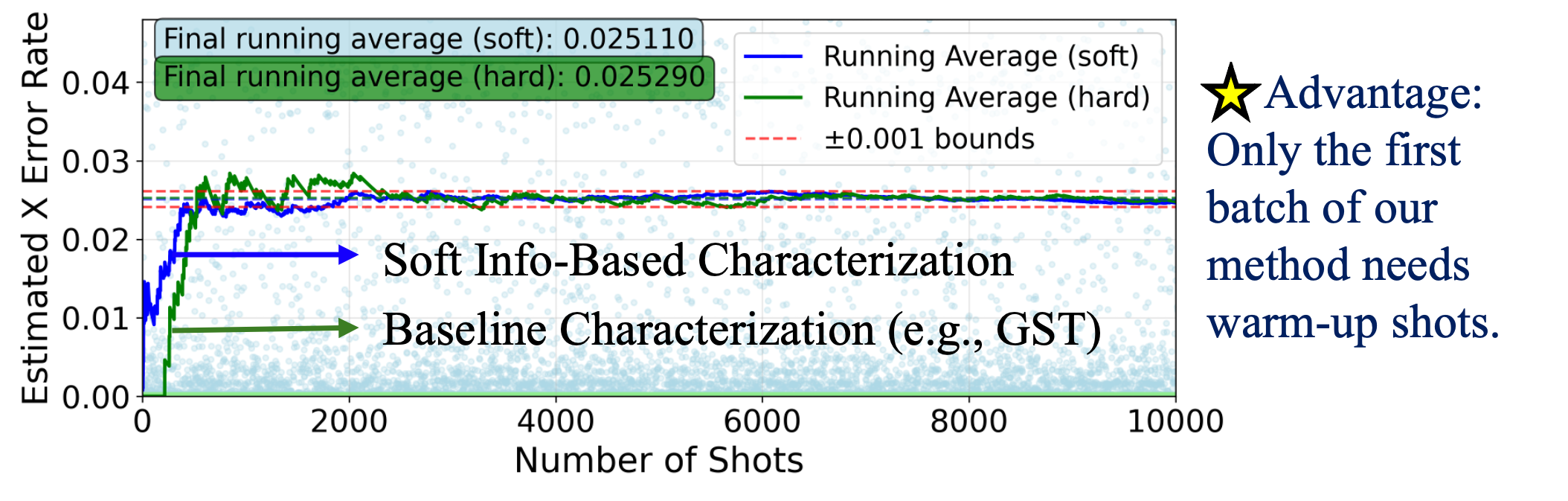}
    \caption{Comparison of logical error rate estimation convergence of baseline characterization method (e.g., GST) and our soft-information-based method. For our method, only the first batch requires this convergence via warm-up shots. }
    \label{fig:convergence}
\end{figure}

\textbf{Error Rate Estimation Convergence.}
Logical error rate characterization converges with more shots. Figure \ref{fig:convergence} compares our method with the baseline method. Note that our method requires warm-up shots to stabilize the logical error estimation only in the first batch of quantum experiments (no prior calibration knowledge). From the second batch, our method can keep updating the estimation with arbitrary circuits as long as it uses the same logical gate set. 
\begin{figure}[!t]
    \centering
    \includegraphics[trim={0 0.5cm 0 0}, width=\columnwidth]{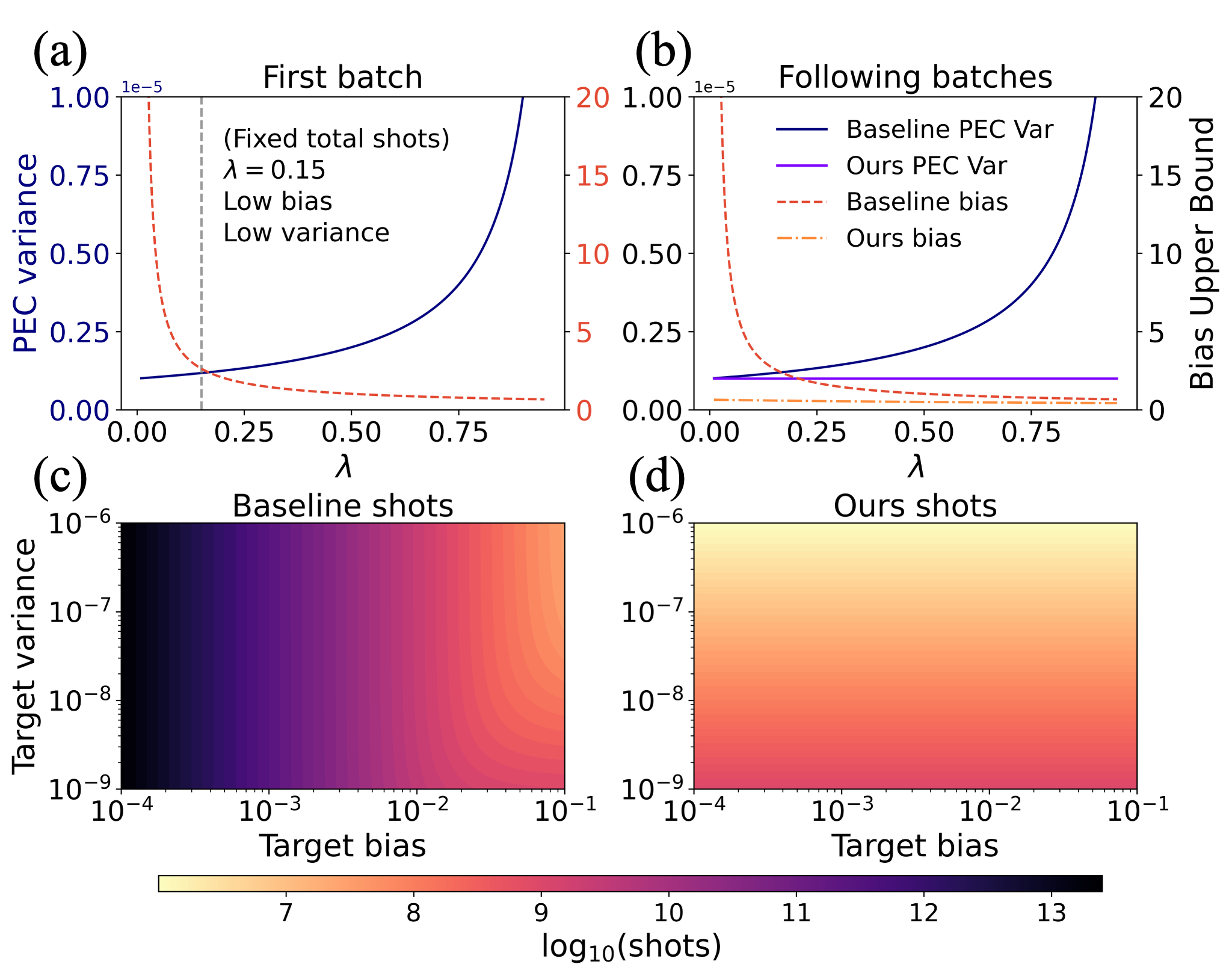}
    \caption{PEC variance and bias tradeoff analysis. In (a) and (b), $\lambda$ denotes the fraction of shots dedicated for characterization. The rest of the shots $(1-\lambda)\times N$ will be used for PEC. (a) Tradeoff at first batch. (b) Tradeoff after first batch, where our method shows significant advantages. In (c) and (d), we present a heatmap of shots required at a given target bias and variance for the baseline and our methods, respectively.
    }
    \label{fig:pec_shot_allocation}
\end{figure}

\textbf{PEC Bias and Variance Analysis}: Figure~\ref{fig:pec_shot_allocation} analyzes in-depth the bias-variance tradeoff in PEC shot allocation, in both the first batch, where warm-up shots are needed, and following batches. After the first batch, our method achieves much lower variance and bias due to the fact that we can use any circuits to update characterization results and all shots to perform PEC.

\begin{figure}[t]
    \centering
    \includegraphics[trim={0 0.6cm 0 0}, width=\columnwidth]{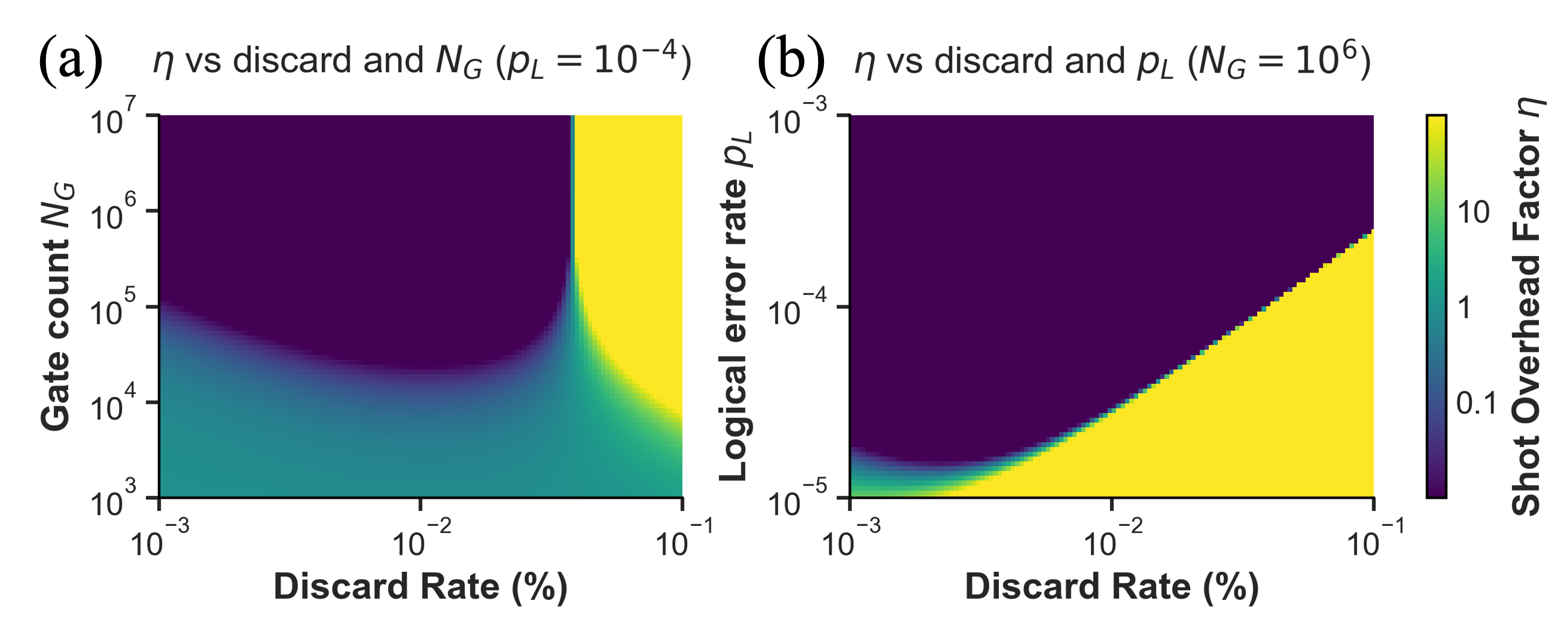}
    \caption{Shot savings of PEC on post-selected logical circuits. $\eta<1$ means shot saving. (a) Heatmap of $\eta$ against discard rate and gate count. (b) Heatmap of $\eta$ against discard rate and original logical error rate.}
    \label{fig:pec_postselected}
\end{figure}

\begin{figure*}[t]
    \centering
    \includegraphics[trim={0 0.8cm 0 0}, width=\textwidth]{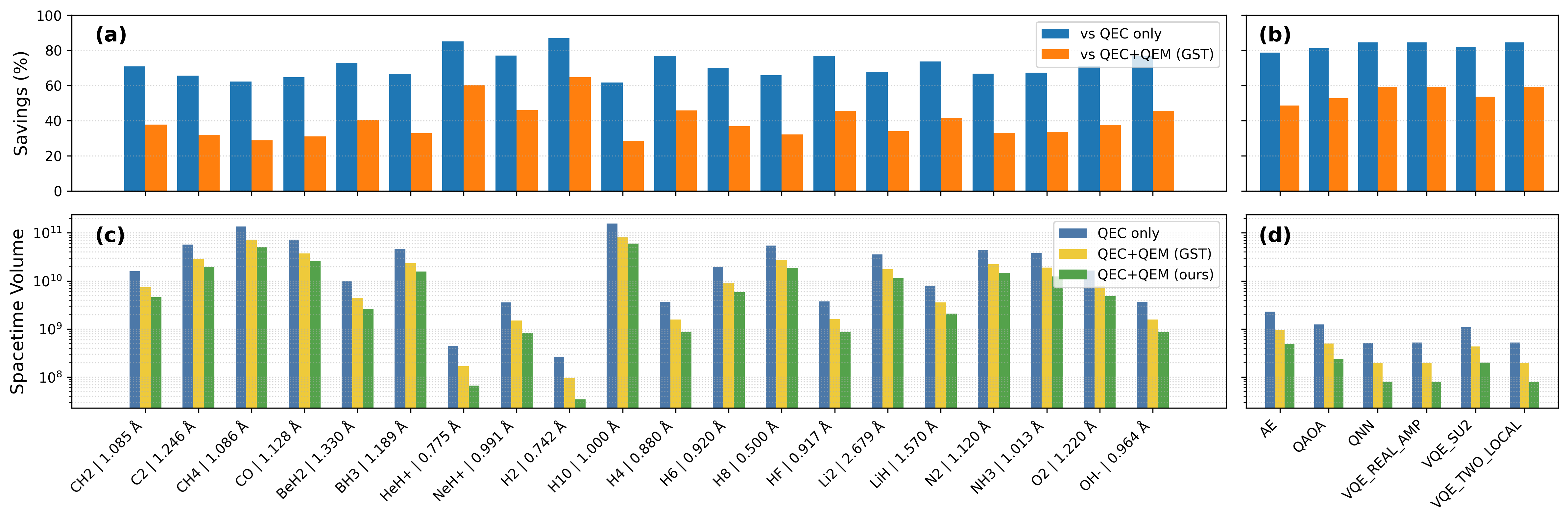}
    \caption{Resource overhead comparison across three architectures for MQT Bench algorithms and PennyLane molecular Hamiltonian 
    simulations. Top row: percentage savings of soft-information-based QEC+QEM (ours) relative to (blue) QEC only and 
    (orange) baseline QEC+QEM for molecular (a) and MQT Bench (b). Bottom panels: absolute spacetime volumes on 
log scale for all three architectures. Our approach achieves $60$-$87\%$ savings versus plain QEC and 
    $30$-$65\%$ savings versus GST-based QEM across all benchmarks.}
    \label{fig:benchmark}
\end{figure*}

\textbf{PEC on Post-Selected Circuits.}
Figure~\ref{fig:pec_postselected} studies numerically the shot overhead saving through performing PEC on post-selected circuits that is motivated in Equation \ref{eq:pec_postselect}. The heatmap demonstrates pronounced savings with low-discard-rate regimes. Note that while our vanilla logical PEC protocol already offers spacetime savings over baseline methods shown in Figure \ref{fig:benchmark}, this PEC on the post-selected channel offers an additional saving on shot overhead.

\subsection{Resource Overhead Benchmarking}
We benchmark the resource overhead reduction with our soft-information-based logical PEC protocol compared to (1) QEC only, and (2) Baseline QEC+QEM with GST for characterization. The characterization cost savings also apply to logical ZNE protocols. To enable a fair and general comparison, we use spacetime-volume as the key overhead metric. Since QEM techniques are mostly used for expectation-value-related experiments, we adopt algorithmic circuits with expectation value outputs from MQT Bench \cite{quetschlich2023mqtbench}  and \textit{chemistry ansätze} from PennyLane \cite{Utkarsh2023Chemistry} for Hamiltonian simulation tasks. 
Specifically, we compile each algorithm with standard Clifford+$T$ gate set that can be efficiently implemented by the surface code \cite{Litinski2019gamesurfaceLS,Suzuki2022qecqem}. Then we extract gate counts $N_G$ and compute the required spacetime volume: the product of qubit count, circuit depth, shot number, and PEC shot overhead factor. 
We compare it for: (A) QEC only, (B) QEC+QEM with GST, and (C) our QEC+QEM with soft information (assume no characterization costs). Here, the PEC shot overhead factor $\gamma=\sum_i|\eta_i|$ and GST shot overhead are assigned according to baseline \cite{Suzuki2022qecqem}. Code distances follow a standard scaling: 
\begin{equation}
    d(\epsilon,N_G) = \frac{2\log((\epsilon/2)/(N_G C_1))}{\log(C_2p/p_{\text{th}})}
\end{equation}
where $C_1=0.03$, $C_2=1$, $p/p_{\text{thres}}=0.85$, and target logical failure $\epsilon=10^{-8}$.

For each algorithm, we report in Figure. \ref{fig:benchmark} total spacetime volumes $(V_A, V_B,V_C)$ on a log scale, together with percentage savings $1-V_C/V_A$ and $1-V_C/V_B$. Across 20 molecular systems (left) and 6 algorithmic circuits 
(right), our approach achieves $60$-$87\%$ savings versus plain QEC (average $73\%$) and $30$-$65\%$ versus GST-based QEM (average $48\%$) across all benchmarks.  \\

\begin{figure}[t]
    \centering
    \includegraphics[trim={0 0.7cm 0 0}, width=\linewidth]{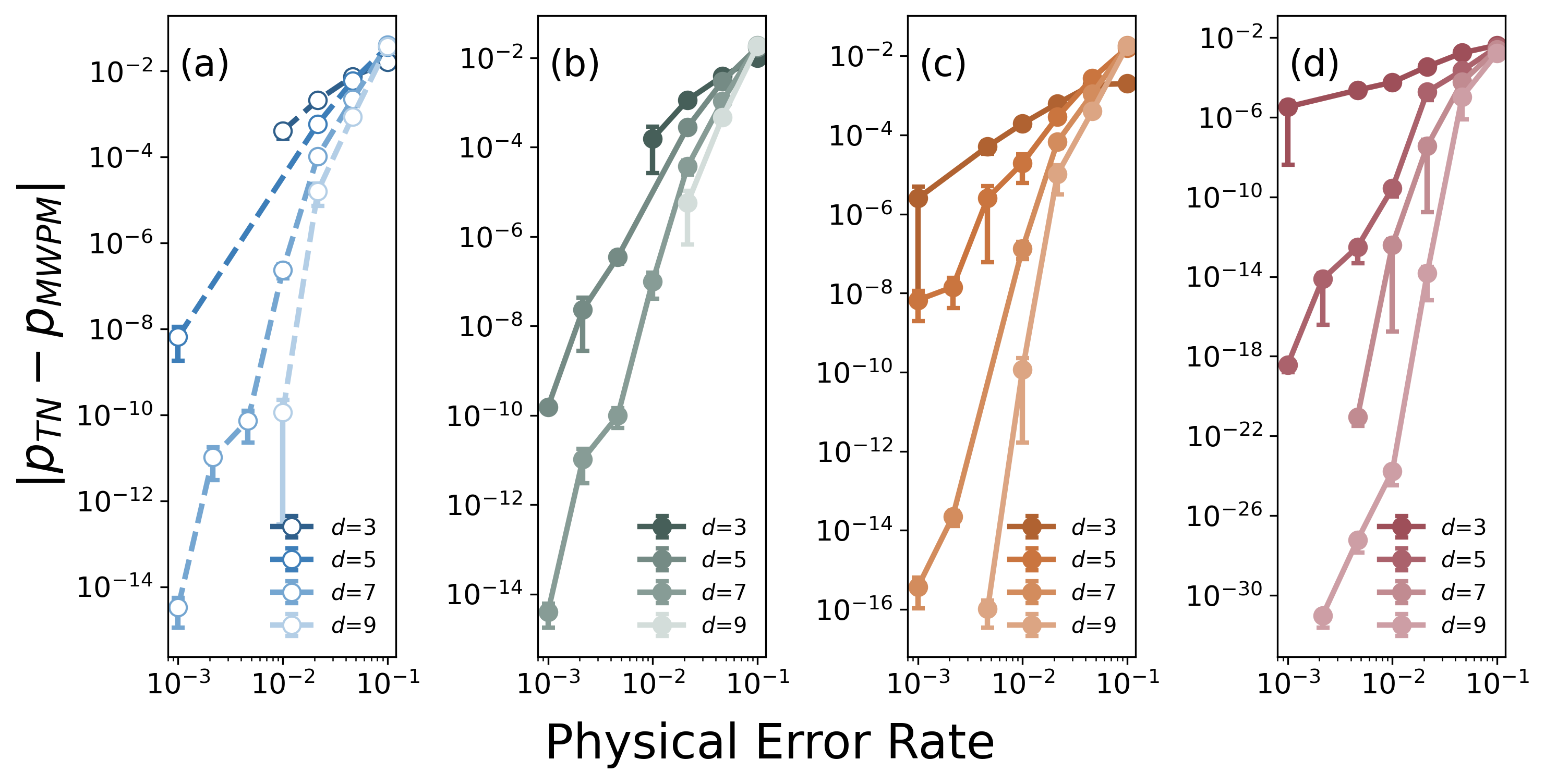}
    \caption{Bias analysis between exact TN and approximate MWPM soft information estimation. Each subplot shows 
    bias versus physical error rate for logical equivalence classes (a) $I$, (b) $X$, (c) $Z$, 
    and (d) $Y$ across code distances $d \in \{3, 5, 7, 9\}$.
    Hollow circles with dashed lines indicate negative bias, while filled circles indicate positive bias.}
    \label{fig:bias_plot}
\end{figure}

\subsection{Soft Information Bias Analysis}
Finally, we numerically evaluate the absolute bias of the MWPM-based soft information calculation compared to the TN-based method in Figure \ref{fig:bias_plot}. Absolute bias $p_{\text{TN}}-p_{\text{MWPM}}$ is the relevant metric when converted to the bias upper bound of PEC and ZNE \cite{Kandala_2019QEMextends}. 
The results show that MWPM overestimates $p_I$ 
and underestimates $p_X$, $p_Y$, and $p_Z$. This occurs because MWPM identifies only the single most-likely error pattern in each class, missing the probability 
mass of degenerate error configurations that contribute to the erroneous states. Notably, the absolute bias decreases exponentially with code distance, making the MWPM method sufficient for high distance and low physical error regime.
\section{Related Works}
\label{sec:related_works}
\emph{Logical QEM}: Logical QEM has been established previously \cite{Suzuki2022qecqem, zhang2025demonstratingquantumerrormitigation, Wahl2023ZNEscaling,Dutkiewicz2025QEMEFT}, but existing studies assume the use of standalone logical characterization protocols. To our knowledge, this is the first work to use decoder soft information for efficient characterization and mitigation of logical errors.

\emph{Logical Characterization}: Logical learning protocols include GST \cite{Brieger2023CGST, Nielsen_2021GST, Suzuki2022qecqem}, randomized benchmarking \cite{Hines2023RBdemo}, and cycle benchmarking \cite{Zhang2025cb}. Our protocol reuses syndrome information from the decoder, requiring no additional quantum runtime.

\emph{QEC Decoder Soft Information}: Decoder soft information has been widely explored in classical decoding but only recently in QEC. Its primary application has been in concatenated code decoding and lattice surgery \cite{gidney2023yokedsurfacecodes, meister2024efficientsoftoutputdecoderssurface, akahoshi2025runtimereductionLS, toshio2025decswitching}.

\emph{Decoder}: Recent developments in decoding algorithms and implementations improve both speed and accuracy \cite{ravi2023btwdecoder, Bausch2023MLdecoding, Tan2023scalabledecoders, maurer2025realtimedecodinggrosscode, chubb2021generaltensornetworkdecoding, viszlai2024predictivewindowdecodingfaulttolerant}. Our method is complementary and decoder-agnostic, applying to any decoder and circuit type. Since soft information calculations are performed offline, there are no stringent latency requirements.
\section{Conclusion}
\label{sec:conlusion}

Through detailed analysis and evaluations, we have shown that decoder soft information is a powerful, intrinsically available resource for logical-level QEM. By turning per-shot, per-gate posterior logical error probabilities from soft-output decoders into unbiased, Rao--Blackwellized estimators of logical channels, our framework enables \textit{in situ} characterization without any additional overhead. Building on this, we introduced three QEM techniques---soft information-based post-selection and runtime abort, efficient logical PEC, and logical ZNE---that reduce logical error rates by more than two orders of magnitude while discarding fewer than 0.1\% of shots, and cut the spacetime overhead by 60--87\% relative to QEC-only and 30--65\% relative to GST-based QEC+QEM across benchmark circuits.

Our results suggest that QEM will remain valuable well into the QEC and FTQC era as an efficient way to use information already produced by QEC decoders and push logical performance beyond what QEC alone can deliver. Future work includes extending this framework to other code families and decoders and incorporating realistic noise models (for instance, the circuit-level noise), enabling architectures where QEC and QEM are co-designed around decoder soft information.

\section{Acknowledgement}
\label{sec:acknowledgement}
The authors would like to thank Yue Wu, Kun Liu, Dantong Li, and Shifan Xu for useful discussions. This work is supported in part by the National Science Foundation (under award CCF-2338063), in part by the Department of Energy Co-Design Center for Quantum Advantage (C2QA), in part by DARPA (HR-0011-23-3-00019), and in part by QuantumCT (under NSF Engines award ITE-2302908). Y.D. acknowledges partial support by Boehringer Ingelheim, and NSF NQVL-ERASE (under award OSI-2435244). External interest disclosure: Y.D. is a scientific advisor to, and receives consulting fees from Quantum Circuits, Inc. A.K. acknowledges support from the NSF (QLCI, Award No. OMA-2120757), IARPA and the Army Research Office (ELQ Program, Cooperative Agreement No. W911NF-23-2-0219).
% ---- bibliography ----
\bibliographystyle{apsrev4-2}
\bibliography{refs}

\end{document}